\documentclass[fleqn,usenatbib]{mnras}

\usepackage{newtxtext,newtxmath}

\usepackage[T1]{fontenc}
\usepackage{hyperref}
\usepackage{color,soul}

\DeclareRobustCommand{\VAN}[3]{#2}
\let\VANthebibliography\thebibliography
\def\thebibliography{\DeclareRobustCommand{\VAN}[3]{##3}\VANthebibliography}

\usepackage{graphicx}	
\usepackage{amsmath}	
\usepackage{tabularx}
\usepackage{lipsum}

\title[PSB properties of post-mergers]{Post-Starburst Properties of Post-Merger Galaxies}

\author[Li et al.]{
Wenhao Li,$^{1}$\thanks{E-mail: wli58@crimson.ua.edu}
Preethi Nair,$^1$
Kate Rowlands,$^2$
Karen Masters,$^3$
David Stark,$^2$
Niv Drory,$^4$
\newauthor
Sara Ellison,$^5$
Jimmy Irwin,$^1$
Shobita Satyapal,$^6$
Amy Jones,$^2$
\newauthor
William Keel,$^1$
Kavya Mukundan$^1$ 
and Zachary Tu$^7$
\\
$^1$Dept. of Physics and Astronomy, The University of Alabama, Tuscaloosa, AL 35487, USA\\
$^2$Space Telescope Science Institute, Baltimore, MD 21218, USA\\
$^3$Dept. of Physics and Astronomy, Haverford College, Haverford, PA 19041, USA\\
$^4$Dept. of Astronomy, The University of Texas at Austin, Austin, TX 78712, USA\\
$^5$Dept. of Physics and Astronomy, The University of Victoria, Victoria, BC V8P 5C2, Canada\\
$^6$Dept. of Physics and Astronomy, George Mason University, Fairfax, VA 22030, USA\\
$^7$Dept. of Astronomy, The University of Washington, Seattle, WA 98105, USA\\
}

\date{Accepted May 12 2023. Received March 24 2023}

\pubyear{2023}

\begin{document}
\label{firstpage}
\pagerange{\pageref{firstpage}--\pageref{lastpage}}
\maketitle

\begin{abstract}
Post-starburst galaxies (PSBs) are transition galaxies showing evidence of recent rapid star formation quenching. 
To understand the role of galaxy mergers in triggering quenching, we investigate the incidence of PSBs and resolved PSB properties in post-merger galaxies using both SDSS single-fiber spectra and MaNGA resolved IFU spectra. 
We find post-mergers have a PSB excess of 10 -- 20 times that relative to their control galaxies using single-fiber PSB diagnostics. 
A similar excess of $\sim$ 19 times is also found in the fraction of central (C)PSBs and ring-like (R)PSBs in post-mergers using the resolved PSB diagnostic. However, 60\% of the CPSBs + RPSBs in both post-mergers and control galaxies are missed by the single-fiber data. By visually inspecting the resolved PSB distribution, we find that the fraction of outside-in quenching is 7 times higher than inside-out quenching in PSBs in post-mergers while PSBs in control galaxies do not show large differences in these quenching directions. In addition, we find a marginal deficit of HI gas in PSBs relative to non-PSBs in post-mergers using the MaNGA-HI data.
The excesses of PSBs in post-mergers suggest that mergers play an important role in triggering quenching. Resolved IFU spectra are important to recover the PSBs missed by single-fiber spectra. 
The excess of outside-in quenching relative to inside-out quenching in post-mergers suggests that AGN are not the dominant quenching mechanism in these galaxies, but that processes from the disk (gas inflows/consumption and stellar feedback) play a more important role.
\end{abstract}

\begin{keywords}
galaxies: evolution -- galaxies: interactions -- galaxies: starburst
\end{keywords}



\section{Introduction}
The galaxy population in the local universe can be broadly divided into two main categories; blue star-forming spirals and red quiescent ellipticals. This bimodality is well characterized by the color vs magnitude diagram of the galaxy population \citep{2003ApJ...594..186B, 2004ApJ...600..681B, 2007ApJS..173..293W, 2014ApJ...787...63J}, which shows the star-forming `blue cloud' separated from the quiescent `red sequence'. This bimodal trend is built up over cosmic time showing star-forming galaxies are quenching their star formation and are evolving from the `blue cloud' to the `red sequence' \citep{2004ApJ...608..752B, 2007A&A...476..137A, 2007ApJ...665..265F, 2007AAS...20916111J}. However, what triggers this evolution and the truncation of star formation is not fully understood. 

Theoretically, galaxy mergers offer a scenario where star formation quenching is expected to be seen. Galaxy mergers can trigger both enhanced star formation \citep{1996ApJ...471..115B, 2006MNRAS.373.1013C, 2007ApJ...660L..51L} and active galactic nuclei (AGN, \citealp{2005Natur.433..604D, 2006ApJS..163....1H,2008ApJS..175..356H, 2009ApJ...696..891H}). Both gas inflows and gas consumption by star formation during mergers \citep{2015MNRAS.448.1107M} and the merger-driven AGN or stellar feedback \citep{2006MNRAS.370..645B, 2006MNRAS.365...11C, 2012ARA&A..50..455F, 2017ApJ...851...18L} 
can cause star formation quenching.
However, star formation quenching due to gas inflows/consumption and AGN/stellar feedback would lead to different resolved star formation histories. 
On one hand, simulations have shown that galaxy mergers can induce gas inflows towards the center, resulting in star formation suppression in the outskirts and enhanced starburst in the center, followed by a truncation of star formation due to gas consumption \citep{2001ASPC..249..735N, 2015MNRAS.448.1107M}. Alternatively, mergers can also trigger widespread starbursts extending over large scales \citep{2005ASSL..329..143S, 2009ApJ...694L.123K, 2009PASJ...61..481S,2013MNRAS.430.1901H}, leading to rapid consumption of the gas in the outskirts first. Both scenarios can cause star formation quenching from the outskirts and progressively moving inward, leading to an outside-in quenching direction. 
On the other hand, simulations support that AGN feedback can drive kinetic winds, which lead to star formation quenching from the center extending to the outskirts, resulting in an inside-out quenching direction \citep{2021MNRAS.508..219N}. This was also seen in observations (e.g., \citealp{2018RMxAA..54..217S}), although \cite{2022MNRAS.515.1430D} suggests that AGN feedback cannot fully quench star formation in short merger timescales but can have long-term effects on the evolution of galaxies over a few Gyrs.
Winds driven by central stellar feedback could also be a potential cause of inside-out quenching \citep{2013ApJ...770...25A}. 

Many previous works studying quenching have focused on post-starburst galaxies (hereafter, PSBs). PSB galaxies are a rare population of galaxies 
spanning a wide range of color from the `blue cloud' through the `green valley' \citep{2007ApJS..173..293W, 2007ApJS..173..342M, 2009MNRAS.398..735Y} to the `red sequence'. They have undergone a recent starburst followed by a rapid truncation of star formation, which are currently transitioning from star-forming to quiescent. PSB galaxies are classified by their spectra which show the presence of strong Balmer absorption from intermediate-age (A-type) stars and an absence of nebular emission from hot young (O- and B-type) stars \citep{1983ApJ...270....7D, 1987MNRAS.229..423C, 1996ApJ...466..104Z, 1999ApJ...518..576P, 2003MNRAS.346..601G, 2005MNRAS.357..937G, 2014MNRAS.444.3408Y}. As their spectra show a superposition of quiescent galaxy spectra and spectral features of A-type stars, PSB galaxies are also known as `E+A' or `K+A' galaxies. These spectral features suggest that a recent starburst happened $<$1 Gyr ago and there is no ongoing star formation currently, indicating a rapid quenching of star formation. Hence, PSB galaxies are the ideal targets to investigate star formation quenching.

Hydrodynamic simulations show that galaxy mergers can trigger a strong starburst followed by a rapid truncation of star formation, and evolve into a PSB phase and eventually the quiescent phase \citep{2009MNRAS.395..144W, 2011ApJ...741...77S, 2020MNRAS.498.1259Z}. In addition to galaxy mergers, cosmological simulations have also shown that a diversity of mechanisms such as ram pressure stripping, shocks and stellar/AGN outflows can also quench star formation in non-merging galaxies and trigger them into a PSB phase \citep{2019NatAs...3..440P, 2019MNRAS.484.2447D}. However, \cite{2019MNRAS.484.2447D} have found that galaxy mergers at z $\sim$ 0 -- 2 tend to be the most frequent cause of PSB galaxies in simulations.

Observationally, nearly all previous studies on the merger-PSB relation began with a PSB galaxy sample and then investigated their morphologies to constrain the merger fractions. Numerous studies have found a significant fraction (13\% -- 64\%) of PSB galaxies showing disturbed morphologies, which indicates a recent merger or interaction with a nearby companion \citep{1996ApJ...466..104Z, 2004MNRAS.355..713B, 2005MNRAS.357..937G, 2008ApJ...688..945Y, 2009MNRAS.396.1349P, 2016ApJ...827..106A, 2016MNRAS.456.3032P}. Studies comparing the merger fraction in PSB galaxies relative to star-forming control galaxies have found a merger excess of a factor of 2 -- 4, suggesting mergers play an important role in triggering the PSB phase of galaxies \citep{2018MNRAS.477.1708P, 2017A&A...597A.134M, 2021ApJ...919..134S, 2022MNRAS.516.4354W}. 

Studies investigating the PSB fraction in merging galaxies are lacking in the literature. The first and only study so far is presented by \cite{2022MNRAS.517L..92E}. They studied a sample of $\sim$ 500 post-merger galaxies with imaging from the Canada-France Imaging Survey (CFIS\footnote{ \url{https://www.cadc-ccda.hia-iha.nrc-cnrc.gc.ca/en/community/unions/ MegaPipe_CFIS_DR3.html}}) and single-fiber spectra from the Sloan Digital Sky Survey (SDSS,  \citealp{2000AJ....120.1579Y}). They used different PSB classification methods and found a PSB excess of a factor of 30 -- 60 in post-merger remnant galaxies compared to non-merging controls, while there is no excess in close galaxy pairs (or early-stage mergers). Their results highly suggest that galaxy mergers can rapidly quench star formation and this process happens mainly after the coalescence of the two nuclei.
 
However, most previous studies which identified PSB galaxies were based on SDSS single-fiber spectra (e.g. \citealp{2005MNRAS.357..937G, 2016ApJ...827..106A, 2018MNRAS.477.1708P, 2022MNRAS.516.4354W, 2022MNRAS.517L..92E}) and would only capture the fluxes within the 2.5$''$ fiber pointing to the center of the galaxies. The single-fiber diagnostic may only be capable of finding PSB galaxies with star formation quenching in the center and will neglect those with star formation quenching in the outskirts. In fact, PSB galaxies have been found to have a variety of resolved quenching morphologies by integral-field-unit (IFU) spectroscopy \citep{2018MNRAS.480.2544R, 2019MNRAS.490.2347Q,2019MNRAS.489.5709C, 2020ApJ...892..146V, 2021PASP..133g2001F, 2022MNRAS.511.4685X}. Hence, IFU observation will provide a more complete census of PSB galaxies, which also enables us to study their resolved quenching properties. 

Gas content is another way to study star formation quenching. As gas can be consumed in star formation or ionized/expelled by stellar or AGN feedback during mergers, gas suppression is expected to be seen when star-formation quenching happens in late-merger systems \citep{2006ApJS..163....1H, 2012ARA&A..50..455F, 2013MNRAS.430.1901H, 2018AAS...23122204P}. However, previous studies on the gas content in merger systems have found different results. Some have found a deficit of neutral atomic hydrogen (HI) gas in mergers compared to non-merging galaxies \citep{1996AJ....111..655H, 2000MNRAS.318..124G, 2022ApJ...934..114Y}, while others have found an enhancement of HI gas in mergers \citep{2004A&A...422..941C, 2017MNRAS.466.4795J, 2018MNRAS.478.3447E}. In addition, some studies have found no significant difference between the HI gas content in galaxy mergers and non-merging galaxies \citep{2015MNRAS.448..221E, 2018ApJS..237....2Z}. 
One possible reason for these contradictory results is the differences in merger stages in the various samples studied.
Previous studies \citep{2012MNRAS.426..549S, 2013MNRAS.433L..59P, 2015MNRAS.454.1742K, 2018ApJ...868..132P, 2019ApJ...881..119P} have found the star formation rates in close pairs increase with decreasing nuclear separations, suggesting star formation activities are more violent near the coalescence phase. Early mergers with low star formation rates may not have consumed much gas while late stage post-merger remnants may show a gas deficit due to gas consumption, underscoring the importance of merger stage in assessing gas content in mergers compared with non-mergers.

Gas depletion is expected to be seen in PSB galaxies where star formation is being quenched or has been quenched. Surprisingly, previous studies have found a significant amount of cold molecular gas in PSB galaxies regardless of merger status \citep{2015MNRAS.448..258R, 2015ApJ...801....1F,2018ApJ...861..123F, 2016ApJ...827..106A, 2022arXiv220411881B, 2022ApJ...929..154S, 2022MNRAS.517L.126S, 2022ApJ...941...93O}, which leads to a question of how PSB galaxies quench star formation without suppressing the gas reservoir. 
\cite{2023ApJ...942...25F} studies the dense molecular gas traced by HCN/HCO$^+$/HNC~(1-0) in six CO-detected PSB galaxies, which are selected with low H$\alpha$ emission (H$\alpha$ $<$ 3\AA) and strong Lick H$\delta_A$ absorption (H$\delta_A - \sigma$H$\delta_A>$ 4\AA, \citealp{2015ApJ...801....1F}). Their results suggests that although PSB galaxies still contain significant (CO-traced) lower density gas reservoirs, the lack of (HCN/HCO$^+$/HNC-traced) dense gas is the reason of their current quiescence.

In this work, we try to understand the relation between star formation quenching and galaxy mergers by studying the post-starburst properties and HI gas content of a sample of post-merger remnant galaxies. In Section~2, we describe our post-merger and control samples and the observational data. In Section~3, we introduce the PSB classification methods used in our work. In Section~4, we present the results of PSB fractions and the resolved quenching history of our sample and we relate them to the HI gas content. We discuss our results in Section~5 and we summarize this work in Section~6.

\begin{figure}
\begin{center}
\begin{minipage}{0.47\textwidth}
\includegraphics[width=\linewidth]{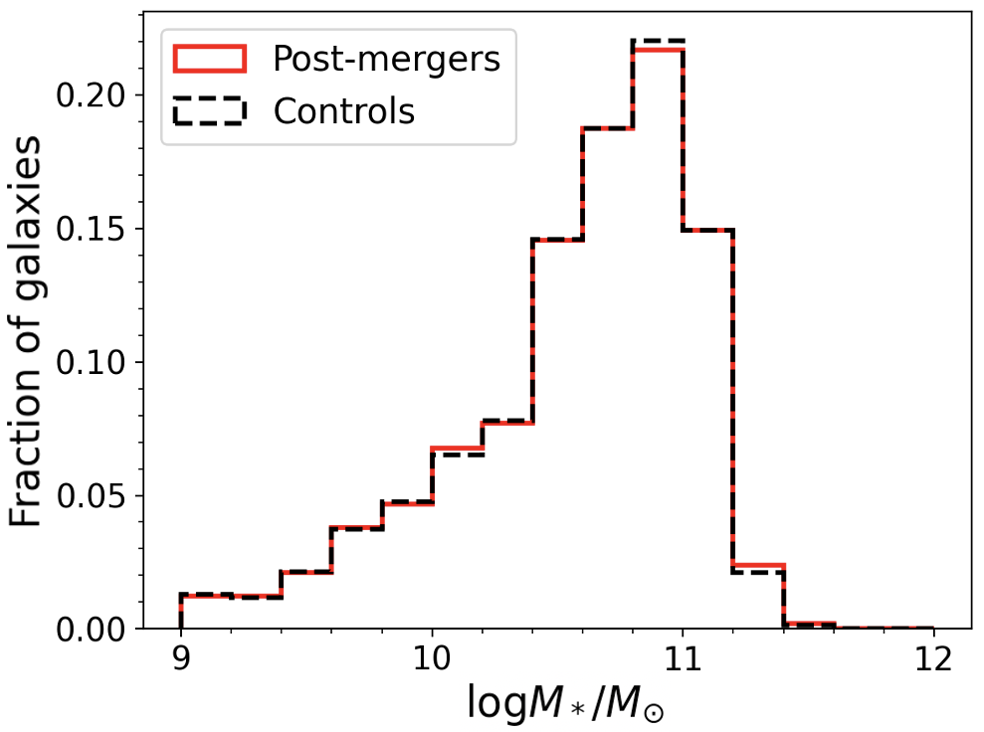}
\end{minipage}
\begin{minipage}{0.47\textwidth}
\includegraphics[width=\linewidth]{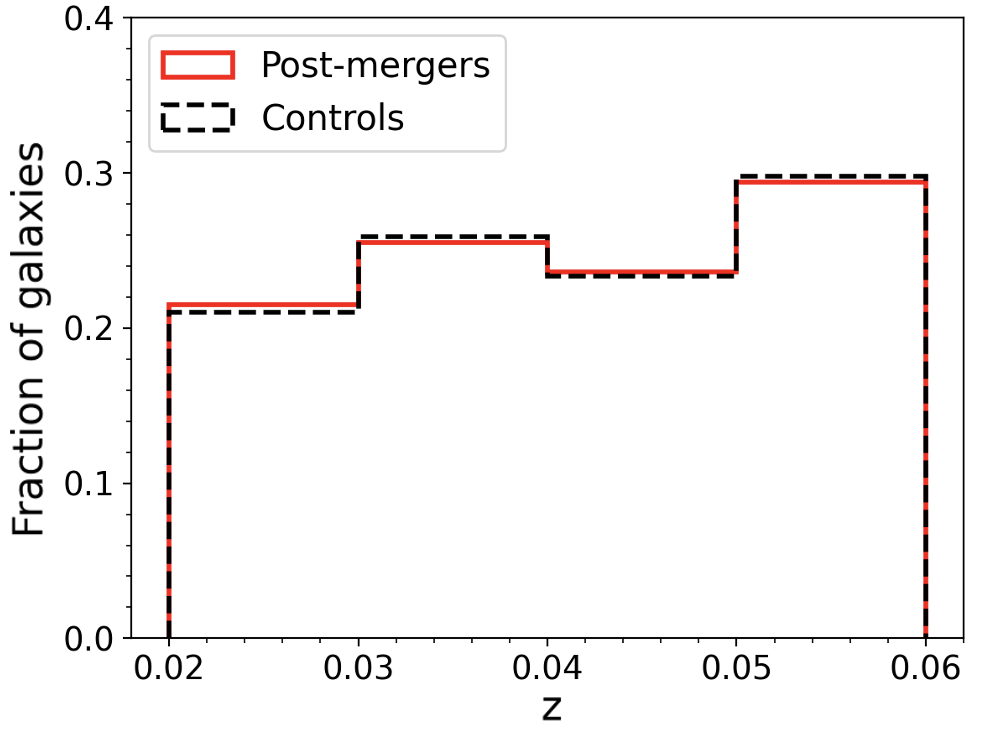}
\end{minipage}
\caption{Stellar mass and redshift distribution of the 1,051 post-merger galaxies (red solid) and the 10,510 control galaxies (black dashed) from SDSS DR14 parent sample. Each post-merger has 10 closest controls matched in both stellar mass and redshift.}
\label{fig:mass_z}
\end{center}
\end{figure}

\begin{figure}
\begin{center}
\begin{minipage}{0.47\textwidth}
\includegraphics[width=\linewidth]{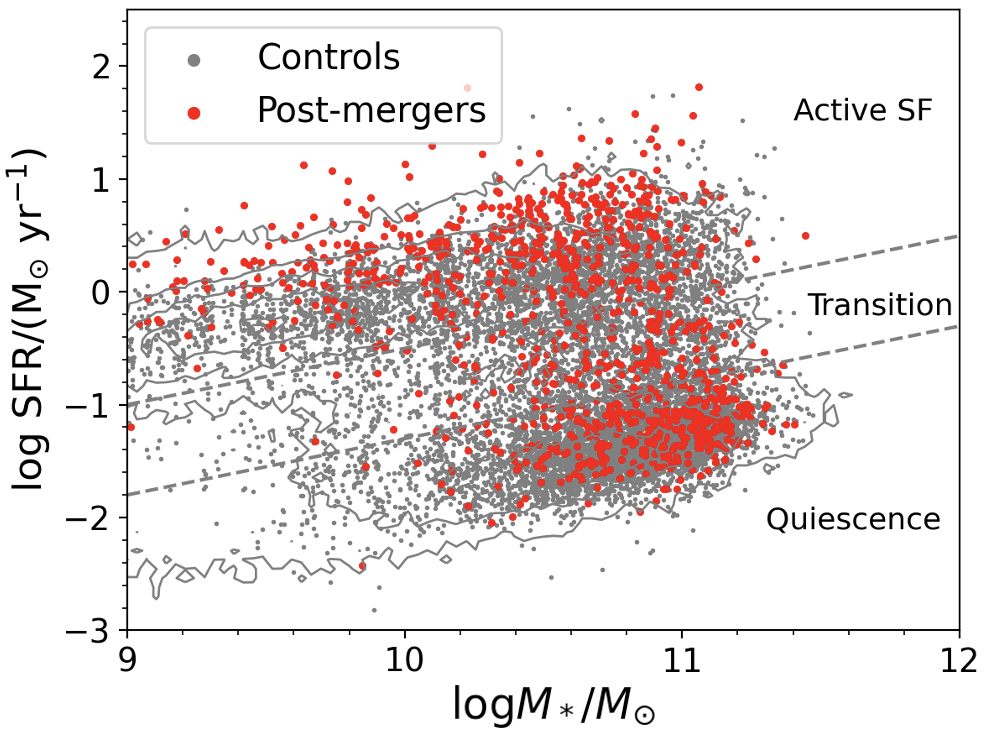}
\end{minipage}
\caption{Star formation rates (SFRs) vs. stellar mass diagram of the DR14 post-merger and control sample. The contours represent the 90,234 galaxies from the DR14 parent sample with the outer most contour including 99\% of the sample. The two division lines are drawn by eye based on the contours, which divide galaxies into active star-forming (SF), transition and quiescence.}
\label{fig:sfr_mass}
\end{center}
\end{figure}

\section{Sample and Data}
\subsection{SDSS Post-Merger Sample}
The parent galaxy sample in this work is a volume-limited ($0.02 \le z \le 0.06$ and 9 $\le$ log M$_*$/M$_{\odot} \le$ 12) morphology catalog of $\sim$113,000 galaxies visually classified from the Sloan Digital Sky Survey (SDSS, \citealp{2000AJ....120.1579Y}) Data Release 14 (Nair et al. 2023, in prep). In short, the $\sim$ 113,000 galaxies in this catalog were visually classified by co-author Nair following the methodology in \cite{2010ApJS..186..427N} using both SDSS imaging and deeper data from the NOAO Legacy Survey \citep{2021AAS...23723503S}. Galaxies were classified into post-merger remnants, double nuclei, close pairs, disturbed galaxies, and non-interacting galaxies. Features such as bars, rings, and spiral arms were also identified. 

In order to provide a robust PSB classification, we required the $r$-band signal-to-noise ratio (S/N) $>$ 10 in the SDSS spectra \citep{2007MNRAS.381..187G,2022MNRAS.516.4354W} and the median S/N $>$ 10 per pixel of the whole spectrum \citep{2019MNRAS.489.5709C}. This reduces the parent sample to 90,234 galaxies. There are 1,165 post-mergers identified in this reduced parent sample. 

For these post-mergers, we built up a control sample of non-interacting galaxies by searching for a unique closest match in stellar mass and redshift within a search range of $\Delta$log M$_*$/M$_{\odot} \pm 0.1$ and $\Delta z \pm 0.005$. We required 10 unique control galaxies for each post-merger to obtain robust statistics. There are 1,051 post-mergers matched with 10 unique controls while the other 114 have less than 10 controls. We only use the 1,051 post-mergers and 10,510 unique control galaxies in the following PSB analysis (DR14 post-merger and control sample, hereafter). 
A Kolmogorov-Smirnov (KS) test provides a p-value of 0.997 for the stellar mass distributions and 0.999 for the redshift distributions between the DR14 post-merger and control sample, confirming that the two samples are indistinguishable in stellar mass and redshift at a 3$\sigma$ confidence level. The stellar mass and redshift distributions of the two samples are shown in Figure~\ref{fig:mass_z}.

The nebular emission and absorption equivalent widths measured by the SDSS single-fiber spectra are obtained from the value-added catalog compiled by the Max Planck Institute for Astrophysics and the John Hopkins University (MPA-JHU, \citealp{2004MNRAS.351.1151B, 2003MNRAS.341...33K, 2004ApJ...613..898T}) DR8. The MPA-JHU catalog also provides derived galaxy properties such as the aperture corrected total stellar masses and star formation rates (SFRs), which are used in this work.  
As described in the MPA-JHU catalog, SFRs are computed within the galaxy fiber aperture using the nebular emission lines \citep{2004MNRAS.351.1151B}. SFRs outside of the fiber are estimated using the galaxy photometry following \cite{2007ApJS..173..267S}. For AGN and galaxies with weak emission lines, SFRs are estimated from the photometry.
All galaxies in our post-merger sample and control sample have measurements in the MPA-JHU DR8 catalog.
As shown in Figure~\ref{fig:sfr_mass}, post-merger galaxies (red points) span a range of star formation rates and stellar masses, migrating from the active star-forming ``blue cloud'' to the quiescent ``red sequence''. Despite the fact that post-mergers have shown strong SFR enhancements (e.g. \citealp{2013MNRAS.435.3627E, 2022MNRAS.514.3294B}), there are many quenched post-mergers as well.
The environment density used in this work is given in \cite{2006MNRAS.373..469B}, which provides measurements for all SDSS DR7 galaxies. 
There are 827/1051 (78.7\%) post-mergers and 8,435/10,510 (80.3\%) control galaxies that have environment density measurements.

\subsection{MaNGA Data}
The Mapping Nearby Galaxies at Apache Point Observatory (MaNGA, \citealp{2015ApJ...798....7B}) is one of the three core programs in the fourth-generation Sloan Digital Sky Survey (SDSS-IV, \citealp{2017AJ....154...28B}). MaNGA is an integral-field-unit (IFU) spectroscopic survey of 10,010 galaxies using the 2.5-meter telescope at the Apache Point Observatory (APO, \citealp{2006AJ....131.2332G}). The IFUs employed in MaNGA are hexagonal fiber bundles with sizes varying from 19 fibers to 127 fibers, which correspond to a diameter from 12$''$ to 32$''$ on the sky. Each fiber has a diameter of 2$''$. Dithered observations are used to cover the field of view \citep{2016AJ....152...83L} with a median spatial resolution of $\sim$ 2.5$''$ FWHM. Spectra are taken using the Baryon Oscillation Spectroscopic Survey (BOSS, \citealp{2013AJ....146...32S}) spectrographs, which covers a wavelength range over 3600-10300 \AA\ at a resolving power of R $\sim$ 2000. MaNGA provides data cubes with a spaxel size of $0.5''$ and reaches a target S/N of 4-8 \AA$^{-1}$ in the outskirts of each MaNGA galaxy \citep{2016AJ....151....8Y}. Two-thirds of the full MaNGA galaxy sample has been observed out to 1.5 effective radius ($R_e$) while the other 1/3 has been observed out to 2.5 $R_e$ \citep{2017AJ....154...86W}. 

The MaNGA Data Analysis Pipeline (DAP, \citealp{2019AJ....158..231W,2019AJ....158..160B}) uses stellar templates drawn from the MILESHC library to determine the stellar kinematics and uses the MASTARSSP library to fit the stellar continuum during the emission-line fitting module. 
We cross-matched the entire parent sample of 90,234 galaxies with the latest MaNGA Product Launch 11 (MPL-11) from SDSS DR17, which corresponds to the DAP version of v3\_1\_1-3.1.0. We found 5,854 galaxies in our sample are observed in MaNGA. Specifically, there are 156 post-mergers observed in MaNGA. 
For these MaNGA observed post-merger galaxies, we also built up a control sample of MaNGA observed non-interacting galaxies, similar to our single-fiber control sample. There are 136 MaNGA observed post-mergers with 10 unique controls within our stellar mass and redshift range (MaNGA observed post-merger and control sample, hereafter). 

Nebular emission and absorption fluxes and equivalent widths are extracted from the MaNGA ``MAPS-SPX-MILESHC-MASTARSSP'' data analysis pipelines, which provide measurements for each individual spaxel. MaNGA also provides the spatially Voronoi-binned (VOR10) data pipelines and the measurements on the binned spectra. However, the measurements on individual spaxels (SPX) provide a better insight of the spaxel distribution. It should be noted that using the Voronoi-binned data pipelines would not affect our results qualitatively.

\begin{figure*}
\begin{center}
\begin{minipage}{0.49\textwidth}
\includegraphics[width=\linewidth]{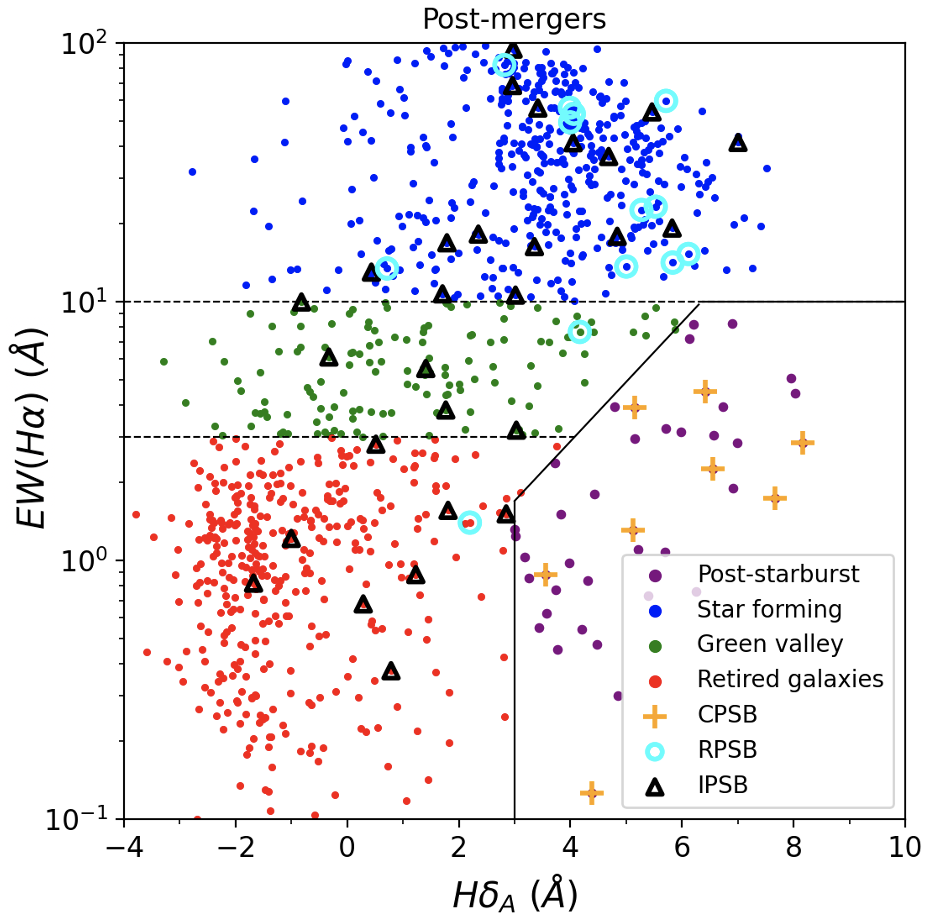}
\end{minipage}
\begin{minipage}{0.49\textwidth}
\includegraphics[width=\linewidth]{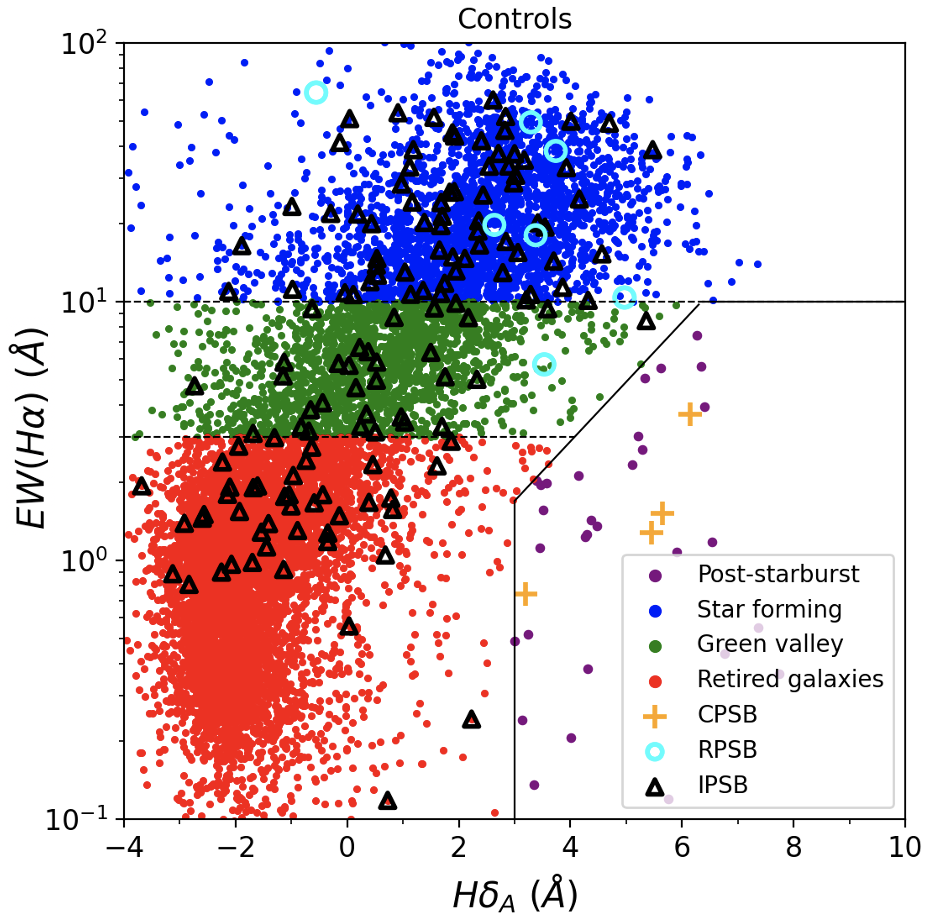}
\end{minipage}
\caption{The equivalent width of H$\alpha$ emission vs. H$\delta_A$ absorption for the DR14 post-merger sample (left) and control sample (right) based on SDSS single-fiber spectra. PSB galaxies (purple) are identified to be located inside the lower right solid box region \citep{2019MNRAS.489.5709C}. Non-PSB galaxies are classified into star-forming (blue dots), green valley (green dots) and retired galaxies (red dots) based on their H$\alpha$ equivalent width. The MaNGA resolved PSBs (Central PSBs, Ring-like PSBs and Irregular PSBs) are over-plotted in different symbols, which will be discussed in Section~\ref{resolvedPSB}.}
\label{fig:psb_classification}
\end{center}
\end{figure*}

\subsection{HI Data}
MaNGA-HI \citep{2019MNRAS.488.3396M} is an HI follow-up program of the SDSS-IV MaNGA Survey galaxies using the Robert C. Byrd Green Bank Telescope (GBT) with supplemented data from the Arecibo Legacy Fast ALFA (ALFALFA, \citealp{2005AJ....130.2598G}) Survey. The HI single-dish observations from GBT have a beam size of FWHM of 9.1$'$ while the observations from ALFALFA have a beam size of FWHM of 3.5$'$. 
The MaNGA HI program selects MaNGA galaxies with z$<$0.05 that lack HI data from ALFALFA and is agnostic to morphological properties. All targets were observed using standard position-switching to the same depth ($\sim$1.5 mJy, $\sim$15 minutes on-source). The detailed observation strategy is described in \cite{2019MNRAS.488.3396M}. 
The HI data used in this work are obtained from the MaNGA-HI DR3 catalog \citep{2021MNRAS.503.1345S}, which provides HI mass measurements for detections and HI mass upper limits for non-detections of $\sim$ 6,000 MaNGA galaxies. 
Following \cite{2021MNRAS.503.1345S}, we excluded sources with contamination from OFF detections, baseline structure, or nearby companions by placing a cut on the source confusion probability p $<$ 0.1.
Out of the 136 post-mergers and 1,360 control galaxies observed in MaNGA, 79 post-mergers and 893 control galaxies have data in the MaNGA-HI DR3 catalog. 
Out of the 79 post-mergers, 37 have HI detections above a 3$\sigma$ threshold and have gas mass measurements. The other 42 are non-detections and only have upper limits on gas mass. In the control sample, 330 control galaxies have HI gas mass measurements while the other 563 have upper limits. We will discuss the HI gas fraction in post-mergers and controls in Section~4.

\section{Post-Starburst Identification}
In this section, we introduce the methods adopted to identify PSB galaxies using the SDSS single-fiber spectra and the MaNGA resolved spectra. 
Post-starburst galaxies can be identified by the presence of strong Balmer (H$\delta$) absorption which indicates an intermediate-age stellar population, and weak or no (H$\alpha$ and/or [OII]) emission suggesting little or no on-going star formation. There are multiple methods which can identify PSB galaxies (see \citealp{2022MNRAS.516.4354W} and references therein).
To compare with previous studies in a consistent way, we first used the traditional `E+A' PSB identification criterion from \cite{2007MNRAS.381..187G} with single-fiber spectra. It requires PSB galaxies to have:
\begin{itemize}
    \item EW(H$\delta_A$) $>$ 5\AA
    \item EW(H$\alpha$) $>$ -3\AA
    \item EW([OII]) $>$ -2.5\AA
    \item $r$-band S/N $>$ 10
\end{itemize}
where the negative equivalent widths indicate emission and positive values indicate absorption. The H$\delta_A$ index is the equivalent width of H$\delta$ absorption feature in the bandpass 4083-4122\AA\ with continuum bandpasses of 4041.6-4079.75\AA\ and 4128.5-4161.0\AA\ \citep{1994ApJS...94..687W, 1997ApJS..111..377W}.
It should be noted that \cite{2012ApJ...761L..16N} show that the PSBs selected by the \cite{2007MNRAS.381..187G} method are not starburst galaxies that are obscured by dust. 
However, this method has strict limits on the EW of H$\alpha$ and [OII] emission lines and may bias against PSB galaxies with shocks or AGN, as well as PSB galaxies that are not yet completely quenched \citep{2011ApJ...737L..38K, 2014ApJ...794L..13A, 2014ApJ...792...84Y}. 

The second method we used on the single-fiber spectra is the \cite{2019MNRAS.489.5709C} PSB selection criteria, which can select PSB galaxies that have quenched rapidly recently and also includes those which have not yet fully quenched their star formation. It requires PSB galaxies to have (see Figure~\ref{fig:psb_classification}):
\begin{itemize}
    \item EW(H$\delta_A$) $>$ 3\AA
    \item EW(H$\alpha$) $>$ -10\AA
    \item log (-EW(H$\alpha$)) $<0.23\ \times$ EW(H$\delta_A$) - 0.46
    \item median spectral S/N $>$ 10 per pixel
\end{itemize}
where the third equation in the \cite{2019MNRAS.489.5709C} selection criteria is determined by an evolutionary track model of a starburst followed by a truncation with an e-folding time of 300 Myr (see details in \citealp{2019MNRAS.489.5709C}). 
For galaxies that are classified as non-PSBs by the \cite{2019MNRAS.489.5709C} method, we classified them into star-forming (EW(H$\alpha$) $<$ -10\AA), green valley (-10\AA\ $<$ EW(H$\alpha$) $<$ -3\AA) and retired galaxies (EW(H$\alpha$) $>$ -3\AA) based on their H$\alpha$ equivalent width (see Figure~\ref{fig:psb_classification}). This is motivated by the evolution track shown in \cite{2019MNRAS.489.5709C} and also by the work of \cite{2011MNRAS.413.1687C}, which classifies star-forming galaxies from retired galaxies based on EW(H$\alpha$). 
As AGN can also ionize gas and contribute to the H$\alpha$ emission, it should be noted that the star-forming and green valley classes identified by this method can be contaminated by AGN. All 1,051 post-mergers and 10,510 control galaxies in our sample satisfy both S/N cuts in these two identification methods.

Unlike the \cite{2007MNRAS.381..187G} method with strict cuts on both H$\alpha$ and [OII] emission lines, the \cite{2019MNRAS.489.5709C} method has loose cuts on H$\alpha$ only and can include more potential PSBs. Hence, we used the \cite{2019MNRAS.489.5709C} selection criteria on the MaNGA resolved spectra to classify each MaNGA spaxel into star-forming, green valley, retired, PSB, and low S/N. To obtain robust measurements, we placed a cut on spaxels with a mean $g$-band spectral S/N $\ge$ 3 per pixel. All bad spaxels (with MaNGA bitmask values $>$ 0) were masked out and were not used in the resolved PSB analysis. In addition to these cuts, we also required a S/N $\ge$ 3 for the EW(H$\delta_A$) to obtain a robust measurement.
We visually inspected the resolved PSB maps and the SDSS cutout images to remove spaxels contaminated by background galaxies or foreground stars.

\begin{table}
	\centering
	\begin{tabular}{|c|c|c|} 
		\hline
		  & \citet{2007MNRAS.381..187G} & \citet{2019MNRAS.489.5709C} \\
		\hline
		1,051 DR14 Post-mergers & 4 (0.4\% $\pm$ 0.2\%) & 41 (3.9\% $\pm$ 0.6\%) \\
        10,510 DR14 Controls & 3 (0.03\% $\pm$ 0.02\%) & 31 (0.3\% $\pm$ 0.1\%) \\
        Excess & 13.1 $\pm$ 9.8 & 13.2 $\pm$ 3.1 \\
        \hline
        136 MaNGA Post-mergers & 1 (0.7\% $\pm$ 0.7\%) & 8 (5.9\% $\pm$ 2.0\%) \\
        1,360 MaNGA Controls & 1 (0.07\% $\pm$ 0.07\%) & 4 (0.3\% $\pm$ 0.1\%) \\
        Excess & 9.9 $\pm$ 14.0 & 20.0 $\pm$ 12.1 \\ 
		\hline
	\end{tabular}
	\caption{The numbers and fractions of PSB galaxies classified by different methods with single-fiber spectra in the DR14 samples and the MaNGA observed samples. The \citet{2019MNRAS.489.5709C} method classifies more PSBs than the \citet{2007MNRAS.381..187G} method while their PSB excesses are consistent. The PSB fractions and excesses in the DR14 post-merger (control) sample are consistent with those in the MaNGA observed post-merger (control) sample.}
	\label{tab:f_psb}
\end{table}

\begin{table*}
	\centering
	\begin{tabular}{|c|c|c|c|c|c|}
		\hline
		   & CPSBs & {RPSBs} & {IPSBs} & {C+R PSBs} & {C+R+I PSBs} \\
		\hline
		136 Post-mergers & 8 (5.9\% $\pm$ 2.0\%) & 13 (9.6\% $\pm$ 2.5\%) & 28 (20.5\% $\pm$ 3.5\%) & 21 (15.4\% $\pm$ 3.1\%) & 49 (36.0\% $\pm$ 4.1\%)\\
        1,360 Controls & 4 (0.3\% $\pm$ 0.1\%) & 7 (0.5\% $\pm$ 0.2\%) & 145 (10.7\% $\pm$ 0.8\%) & 11 (0.8\% $\pm$ 0.2\%) & 156 (11.5\% $\pm$ 0.9\%)\\
        Excess & 20.0 $\pm$ 12.1 & 18.6 $\pm$ 8.5 & 1.9 $\pm$ 0.4 & 19.1 $\pm$ 6.9 & 3.1 $\pm$ 0.4\\
		\hline
	\end{tabular}
	\caption{The numbers and fractions of PSB galaxies classified by using the \citet{2019MNRAS.489.5709C} method and MaNGA resolved spectra. PSB galaxies are classified into central (C) PSBs, ring-like (R) PSBs and irregular (I) PSBs. The last two columns show the C+R PSBs only and the total (C+R+I) resolved PSBs, respectively.}
	\label{tab:cri_psb}
\end{table*}

\section{Results}
\subsection{Post-Starburst Galaxies from SDSS Single-Fiber Spectra}
To classify PSB galaxies using the SDSS single-fiber spectra, we obtained the equivalent widths of H$\alpha$, [OII] and H$\delta_A$ from the MPA-JHU catalog. 

Using the traditional \cite{2007MNRAS.381..187G} method, we identified four PSB galaxies from the DR14 post-merger sample and three PSB galaxies from the DR14 control sample. The PSB fraction is 0.4\% $\pm$ 0.2\% \footnote{The binomial 1$\sigma$ error is calculated as $\sqrt{f (1-f)/N}$, where f is the fraction and N is the sample size.} in post-mergers and 0.03\% $\pm$ 0.02\% in control galaxies, which implies a PSB excess of a factor of 13.1 $\pm$ 9.8 in the DR14 post-merger sample.

Using the \cite{2019MNRAS.489.5709C} method with the DR14 samples, we classified the 1,051 post-mergers into 479 (45.6\% $\pm$ 1.5\%) star-forming, 142 (13.5\% $\pm$ 1.1\%) green valley, 389 (37.0\% $\pm$ 1.5\%) retired and 41 (3.9\% $\pm$ 0.6\%) PSB galaxies (see Figure~\ref{fig:psb_classification}). The 10,510 control galaxies are classified into 2535 (24.1\% $\pm$ 0.4\%) star-forming, 1869 (17.8\% $\pm$ 0.4\%) green valley, 6075 (57.8\% $\pm$ 0.5\%) retired and 31 (0.3\% $\pm$ 0.1\%) PSB galaxies. It implies a PSB excess of a factor of 13.2 $\pm$ 3.1 in the DR14 post-merger sample. The PSB excess of this identification method is consistent with that in the traditional \cite{2007MNRAS.381..187G} method. Table~\ref{tab:f_psb} shows the PSB fractions and excesses in our samples classified by these two different methods.
Both methods suggest that star formation quenching is more common in post-mergers than non-merging galaxies. A Fisher exact test shows that PSBs and post-mergers are strongly related with a p-value = $10^{-24}$ for null hypothesis. However, the SDSS single-fiber spectra only captures the fluxes within the 2.5$''$ fiber, which covers mainly the central region of galaxies. Results based on single-fiber spectra may miss PSB galaxies where star formation quenching happens not in the center but in the outer disk or outskirts. In the following section, we investigated the PSB fraction in our samples using the MaNGA resolved spectra.

\begin{figure}
\begin{center}
\begin{minipage}{0.47\textwidth}
\includegraphics[width=\linewidth]{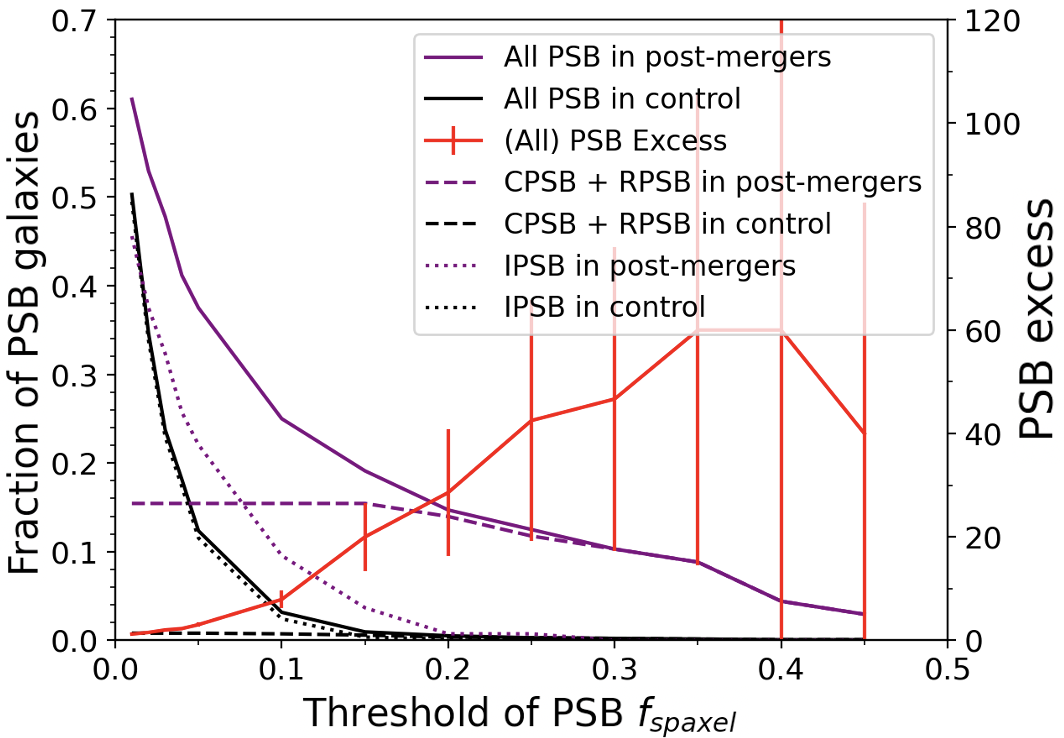}
\end{minipage}
\caption{Fraction of galaxies classified as resolved PSB in post-mergers (purple) and control galaxies (black) vs. the used threshold of PSB spaxel fraction. All three types of PSBs (CPSBs, RPSBs, and IPSBs) in total are shown in solid curves. The red curve with error bars shows the PSB excess in post-mergers relative to controls. The dashed curves show only the CPSB + RPSB and the dotted curves show only the IPSB. There is always a PSB excess in post-mergers relative to control galaxies at any considered thresholds. }
\label{fig:psb_threshold}
\vspace{-0.2cm}
\end{center}
\end{figure}

\begin{figure*}
\begin{center}
\begin{minipage}{0.8\textwidth}
\includegraphics[width=\linewidth]{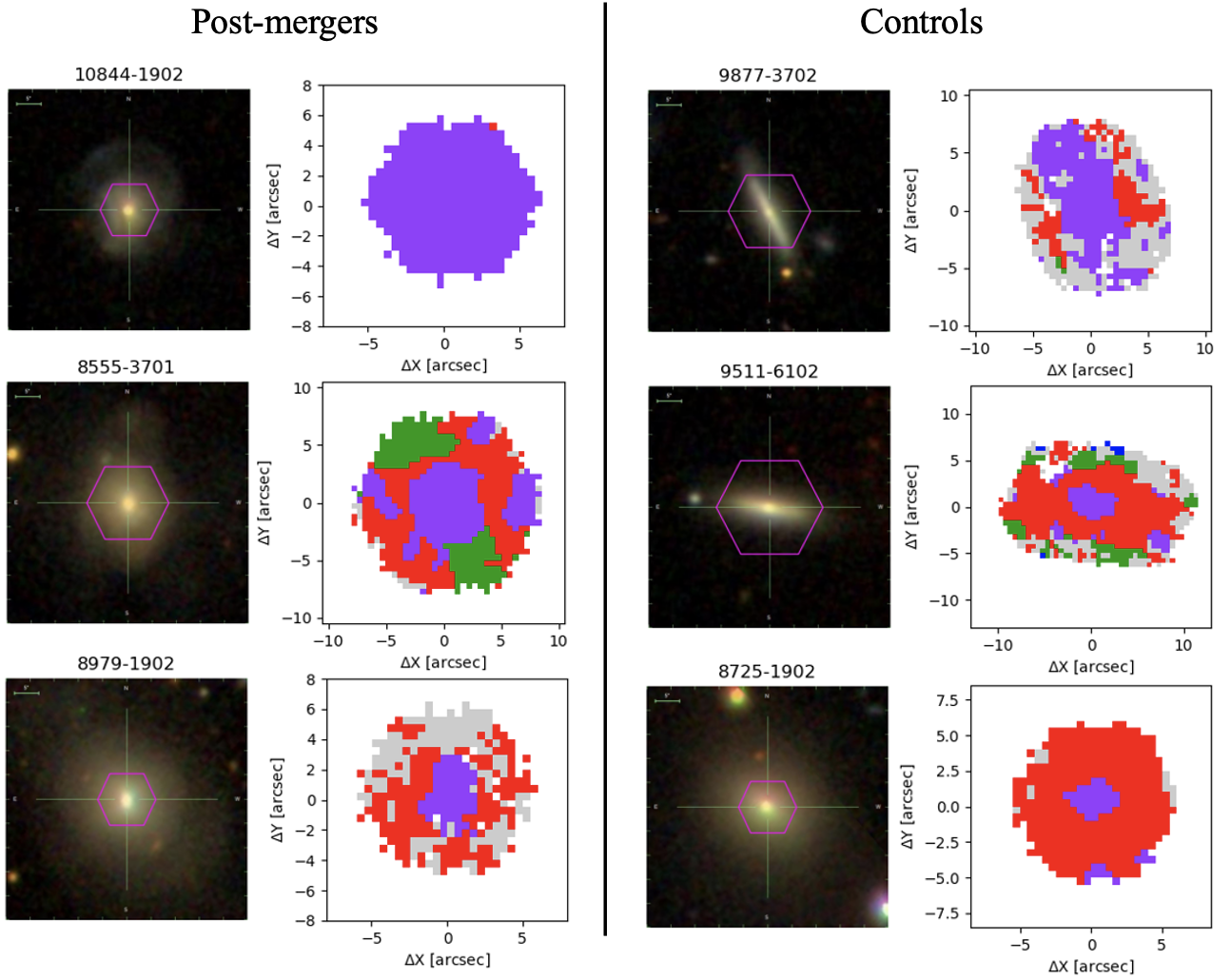}
\end{minipage}
\caption{Examples of SDSS cutout images and resolved PSB classification of CPSBs in post-mergers on the left column and in control galaxies on the right. The coverage of the MaNGA plate is shown in magenta hexagons on the images. Spaxels are classified into star-forming (blue), green valley (green), retired (red), PSB (purple) and low S/N (grey). White color indicates no coverage or bad pixels. }
\label{fig:psb_cutout1}
\vspace{-0.2cm}
\end{center}
\end{figure*}

\begin{figure*}
\begin{center}
\begin{minipage}{0.8\textwidth}
\includegraphics[width=\linewidth]{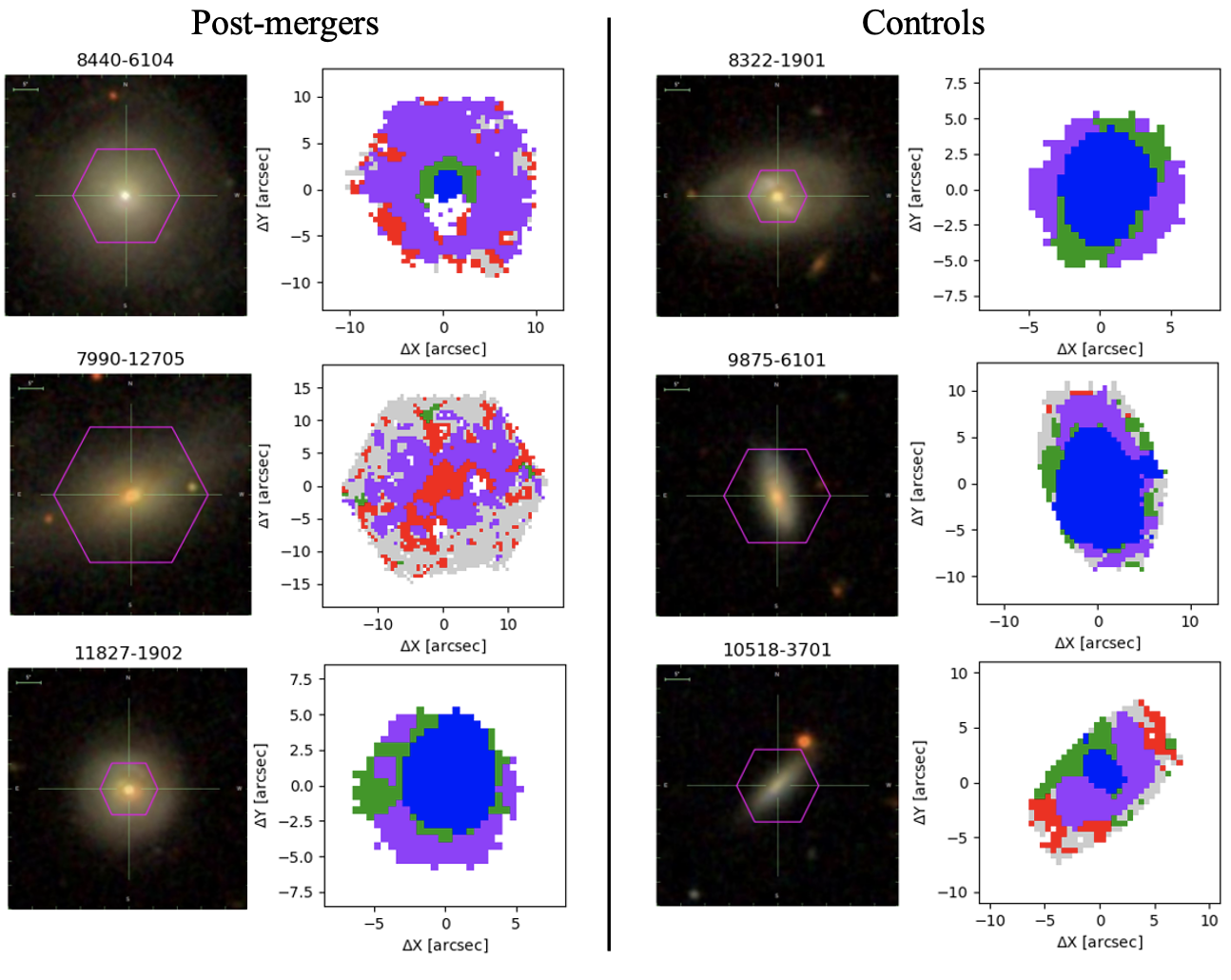}
\end{minipage}
\caption{Examples of SDSS cutout images and resolved PSB classification of RPSBs in post-mergers on the left column and in control galaxies on the right. Colors are the same as in Figure~\ref{fig:psb_cutout1}.}
\label{fig:psb_cutout2}
\vspace{-0.2cm}
\end{center}
\end{figure*}

\begin{figure*}
\begin{center}
\begin{minipage}{0.8\textwidth}
\includegraphics[width=\linewidth]{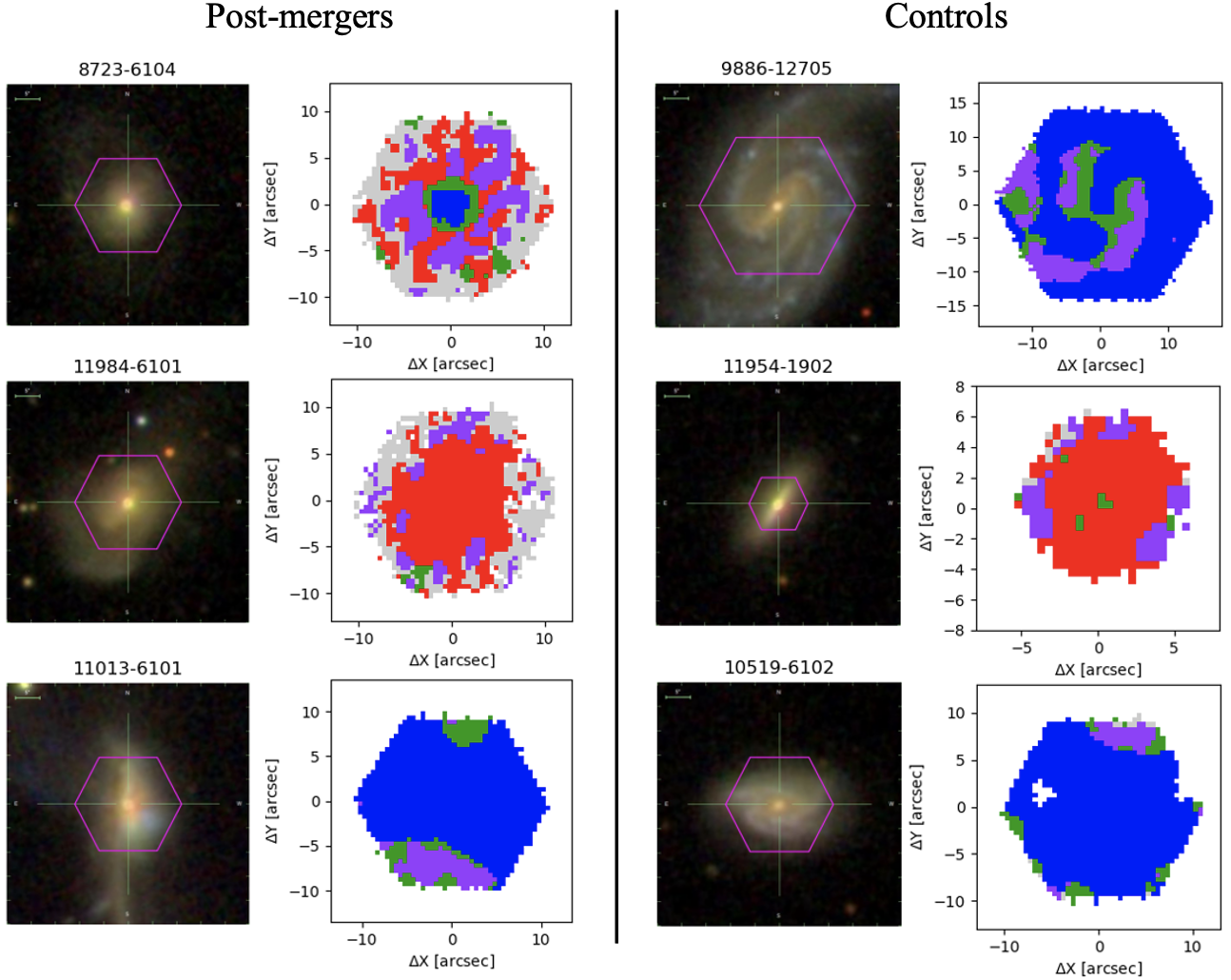}
\end{minipage}
\caption{Examples of SDSS cutout images and resolved PSB classification of IPSBs in post-mergers on the left column and in control galaxies on the right. Colors are the same as in Figure~\ref{fig:psb_cutout1}.}
\label{fig:psb_cutout3}
\end{center}
\end{figure*}

\vspace{0.2cm}
\subsection{Post-Starburst Galaxies from MaNGA IFU Spectra}\label{resolvedPSB}
In this section, we used the MaNGA IFU spectroscopy to present a resolved PSB identification to our MaNGA observed samples. There are 136 post-mergers and 1,360 control galaxies observed in MaNGA. The single-fiber PSB fractions and excesses in these MaNGA observed subsamples are shown in Table~\ref{tab:f_psb}, which are consistent with those in the DR14 post-merger and control sample.

By using the \cite{2019MNRAS.489.5709C} PSB selection criteria, we classified all the spaxels of a galaxy into either one of the following five classes: star-forming, green valley, retired, post-starburst and low S/N. In this way, we obtained a spatially resolved PSB classification map for each galaxy.
To identify a galaxy as a resolved PSB galaxy, we require a minimum fraction of spaxels to be classified as PSB. Figure~\ref{fig:psb_threshold} shows the relation of the fraction of resolved PSBs classified in post-mergers and controls as a function of the threshold applied on the PSB spaxel fraction. 
There is a PSB excess in post-mergers relative to controls at any considered thresholds (see the red line in Figure~\ref{fig:psb_threshold}). The intense drop of the PSB fractions in control galaxies at thresholds $<$ 5\% could be due to non-PSB galaxies with sporadic PSB spaxels, which are misclassified as PSB galaxies.
Large thresholds may overlook potential resolved PSB galaxies and also induce larger errors in statistics. Hence, we choose 5\% as our fiducial threshold for the purposes of quoting statistics.
It should be noted that \cite{2019MNRAS.489.5709C} used a minimum number of six contiguous spaxels to classify PSB galaxies. However, the PSB galaxies classified in our work by using the 5\% threshold all have more than 16 contiguous spaxels. Hence, we have a stricter selection criterion. 

When using a 5\% threshold, 49 (36.0\% $\pm$ 4.1\%) out of the 136 MaNGA observed post-mergers are classified as resolved PSBs while 156 (11.5\% $\pm$ 0.9\%) out of the 1,360 MaNGA observed control galaxies are classified as PSBs. This implies a resolved PSB excess of a factor of 3.1 $\pm$ 0.4 in post-mergers compared to controls. A Fisher exact test shows resolved PSBs and post-mergers are significantly related with a p-value = $10^{-11}$ for the null hypothesis. 

We visually inspected the resolved PSB maps and classified the resolved PSB galaxies into three types: central PSBs (CPSBs), ring-like PSBs (RPSBs) and irregular PSBs (IPSBs).
CPSBs are classified as galaxies showing PSB contiguous spaxels concentrated in the center or all over the galaxies (see Figure~\ref{fig:psb_cutout1}). RPSBs are classified as galaxies showing a full or partial ring of contiguous PSB spaxels (see Figure~\ref{fig:psb_cutout2}). Galaxies showing both central PSB regions and ring PSB regions are still classified as CPSBs. IPSBs are classified as galaxies showing sporadic PSB regions without a central or ring-like concentration (see Figure~\ref{fig:psb_cutout3}). 
Two of the scientists in our research group (Li and Nair) did the visual classification and agreed with each other in all cases. Out of the 49 resolved PSBs in the MaNGA observed post-merger sample, eight are classified as CPSBs, 13 are classified as RPSBs and 28 are IPSBs. Out of the 156 resolved PSBs in the MaNGA observed control sample, four are classified as CPSBs, seven are RPSBs and 145 are IPSBs. 
Table~\ref{tab:cri_psb} shows the fraction and excess of each type of resolved PSBs in the MaNGA observed post-mergers and controls. There are much more CPSBs and RPSBs in post-mergers than in controls, while IPSBs are just slightly more common in post-mergers. This is true regardless of the threshold of PSB spaxel fraction used as can be seen in Figure~\ref{fig:psb_threshold}. 
However, the fractions of IPSBs in post-mergers and controls strongly increases at lower PSB spaxel thresholds (dotted lines) and have roughly the same trend. This suggests that the main difference between the resolved PSB galaxies fraction in post-mergers and control galaxies is due to the CPSBs and RPSBs, which are more common in post-mergers and likely due to true quenching activities. The IPSBs may instead be a signature of sporadic star formation decay rather than permanent quenching. However, some IPSBs show large contiguous PSB regions on the edge of the galaxies or in weird distributions. We cannot rule out that some of these IPSBs are due to true quenching activities.

Figure~\ref{fig:psb_cutout1}, \ref{fig:psb_cutout2} and \ref{fig:psb_cutout3} show examples of the SDSS cutout images and resolved PSB classification maps of CPSBs, RPSBs and IPSBs respectively. PSBs in post-mergers are shown on the left and PSBs in controls are on the right. Spaxels are classified into star-forming (blue), green valley (green), retired (red), PSB (purple) and low S/N (grey). 
The ring regions start outside of the bulge and mostly extend to the edge of the MaNGA footprint. For the eight out of 13 post-merger RPSBs and four out of seven control RPSBs observed in the MaNGA primary sample, their ring regions reach out to $\sim$1.5$R_e$, while the ring regions in the other RPSBs observed in the secondary sample reach out to $\sim$2.5$R_e$.
These three subclasses of resolved PSB galaxies are also over-plotted in the single-fiber classification shown in Figure~\ref{fig:psb_classification}. 
All the CPSBs identified by the resolved spectra are also identified as PSBs by the single-fiber spectra (see the orange crosses in Figure~\ref{fig:psb_classification}). 
Most of the RPSBs (11 out of 13) in post-mergers and (6 out of 7) controls are classified as star-forming in the center by the single-fiber spectra. It suggests quenching only happens in the outer disk but the galaxies are still actively forming stars in the center, indicating an outside-in quenching direction. 
IPSBs span a range in star-forming, green valley and retired.

Using resolved MaNGA IFU data, the PSB fraction in post-mergers increases to 36.0\% compared to 5.9\% in single-fiber spectra while for control galaxies the PSB fraction increases from 0.3\% to 11.5\%. While the overall excess of PSBs in post-mergers decreases from $\sim 20$ in single-fiber to $\sim 3$ with resolved IFU data, the overall PSB fraction increases. 
When considering CPSBs and RPSBs only, the PSB fraction in post-mergers is 15.4\% $\pm$ 3.1\% while the fraction in control galaxies is 0.8\% $\pm$ 0.2\% which implies a PSB excess of a factor of 19.1 $\pm$ 6.9 in post-mergers. 
Although this is consistent with the PSB excess found by the single-fiber spectra (see Table~\ref{tab:f_psb}), the resolved spectra still reveal a lot of PSBs (specifically the RPSBs) which are missed by the single-fiber spectra.
Thus IFU spectroscopy is crucial to provide a more complete census of the resolved PSBs population. However, the origin of IPSBs is unclear. 

\subsubsection{Direction of Star-Formation Quenching}\label{AGNresolvedPSB}
In this section, we discuss the resolved star formation history and the quenching directions in the resolved PSBs identified in the MaNGA observed post-mergers and controls. By visually inspecting the resolved PSB classification maps, we classified our PSB galaxies into six classes with different quenching properties:

\textbf{(i) AGN contaminated}: 
AGN can contribute to the H$\alpha$ emission and contaminate our classification of star-forming or green valley spaxels.  For PSB galaxies showing star-forming/green valley spaxels in the center and hosting an AGN at the same time, we cannot know whether their H$\alpha$ emission is due to star formation or the AGN. Hence, we cannot fully understand their resolved star formation history and the quenching direction. These galaxies will be classified into the `AGN contaminated' class. The optical AGN identification using both single-fiber diagnostic and resolved diagnostic is presented in the Appendix. In particular, the resolved AGN diagnostic is able to reveal the AGN which are obscured by dust and are not detected in the single-fiber diagnostic (see the Appendix).
It should be noted that PSB galaxies showing PSB/retired spaxels in the center are not affected by the AGN signals.

\textbf{(ii) Outside-in quenching}: For PSB galaxies showing PSB/retired spaxels in the outer regions surrounding star-forming/green valley spaxels in the center (and not hosting an AGN), their resolved star formation history suggests quenching happens in the outskirts while star formation is still on-going in the center. This is consistent with an outside-in quenching direction, which is related to processes happening on the disk (e.g., gas inflows or consumption during mergers or tidal stripping). PSB galaxies showing retired spaxels in the outer regions surrounding PSB spaxels in the center also suggests that on-going quenching is happening in the center while the outskirts have already been quenched. This is also consistent with an outside-in quenching direction.

\textbf{(iii) Inside-out quenching}: For PSB galaxies showing PSB/retired spaxels in the center surrounded by star-forming/green valley spaxels in the outer regions, their resolved star formation history is consistent with an inside-out quenching direction, which is likely related to processes in the center (e.g., AGN feedback or central starburst feedback driven by bars/secular processes). PSB galaxies showing retired spaxels in the center surrounded by PSB spaxels in the outer regions also suggests that the star formation in the center has already been quenched but quenching is still happening in the outskirts. This is also consistent with inside-out quenching. As AGN only contaminates spaxels with strong EW(H$\alpha$) (star-forming/green valley), the existence of AGN will not affect our classification of PSB and retired spaxels in the center.

\textbf{(iv) Globally quenching}: Galaxies showing PSB regions all over are quenching their star formation globally. There is no clear quenching direction. We classified these population as `globally quenching'.

\textbf{(v) Quenched overall}: Some galaxies show retired spaxels nearly all over the galaxies with small patches of PSB spaxels randomly distributed. It indicates that star formation has been fully quenched all over the galaxies. These could be the end products of either outside-in or inside-out quenching processes. We classified these galaxies as quenched overall.

\textbf{(vi) No clear trend}: In addition to the five classes mentioned above, other PSB galaxies either show PSB regions stochastically distributed or having irregular distribution. These features may be related to secular processes or sporadic star formation. However, they do not have a clear overall trend like the other five classes, and hence we classified these galaxies as a separate population as `no clear trend'. 

In the 49 PSB galaxies in post-mergers, we identified 11 (22.4\% $\pm$ 6.0\%) as being AGN contaminated, 22 (44.9\% $\pm$ 7.1\%) showing outside-in quenching, three (6.1\% $\pm$ 3.4\%) showing inside-out quenching, three (6.1\% $\pm$ 3.4\%) showing globally quenching, four (8.2\% $\pm$ 3.9\%) being quenched overall, and six (12.2\% $\pm$ 4.7\%) exhibiting no clear trend. Table \ref{tab:f_quench} summarizes the fraction of each quenching direction in post-mergers.
Specifically, the eight CPSB galaxies are classified into five outside-in quenching and three globally quenching. In the 13 RPSB galaxies, we classified three as being AGN contaminated, nine showing outside-in and one showing inside-out quenching signatures. For the 28 IPSB galaxies, eight are AGN contaminated, eight exhibit outside-in quenching, two inside-out, four quenched overall, and six with no clear trend.

\begin{table*}
	\centering
	\begin{tabular}{|c|c|c|c|c|c|c|} 
		\hline
		   & {AGN} & {Outside-in} & {Inside-out} & {Globally quenching}  & {Quenched overall} & {No clear trend} \\
		\hline
		PM & 22.4\% $\pm$ 6.0\% & 44.9\% $\pm$ 7.1\% & 6.1\% $\pm$ 3.4\% & ~6.1\% $\pm$ 3.4\% & 8.2\% $\pm$ 3.9\% & 12.2\% $\pm$ 4.7\% \\
        Control & 17.9\% $\pm$ 3.1\% & 19.2\% $\pm$ 3.2\% & 25.6\% $\pm$ 3.5\% & 0.6\% $\pm$ 0.6\% & 3.2\% $\pm$ 1.4\% & 33.3\% $\pm$ 3.8\% \\
        Excess & 1.3 $\pm$ 0.4 & 2.9 $\pm$ 0.6 & 0.2 $\pm$ 0.1 & 10.2 $\pm$ 11.6 & 2.6 $\pm$ 1.7 & 0.4 $\pm$ 0.1 \\
		\hline
	\end{tabular}
	\caption{The fractions and excesses of different quenching situation in 49 PSBs in post-mergers and 156 PSBs in controls. Outside-in quenching is more common in post-mergers while inside-out quenching and quenching with no clear trend are more common in non-merging controls. Fractions of galaxies which are globally quenching, being quenched overall, or being AGN contaminated are similar in post-mergers and control galaxies considering errors.}
	\label{tab:f_quench}
\end{table*}

\begin{table*}
	\centering
	\begin{tabular}{|c|c|c|c|c|} 
		\hline
		   & {AGN} & {Outside-in} & {Inside-out} & {Globally quenching}\\
		\hline
		PM & 14.3\% $\pm$ 7.6\% & 66.7\% $\pm$ 10.3\% & 4.8\% $\pm$ 4.7\% & 14.3\% $\pm$ 7.6\% \\
        Control & 9.1\% $\pm$ 8.7\% & 81.8\% $\pm$ 11.6\% & -- & 9.1\% $\pm$ 8.7\% \\
        Excess & 1.6 $\pm$ 1.7 & 0.8 $\pm$ 0.2 & -- & 1.6 $\pm$ 1.7 \\
		\hline
	\end{tabular}
	\caption{The fractions and excesses of different quenching situation in CPSBs + RPSBs in post-mergers (N = 21) and controls (N = 13). Once we restrict our PSBs to CPSBs and RPSBs only, the fractions of each quenching direction in post-mergers is similar to that in controls.}
	\label{tab:f_quench_CR}
\end{table*}

In the 156 PSB non-merging controls, we identified 28 (17.9\% $\pm$ 3.1\%) being AGN contaminated, 30 (19.2\% $\pm$ 3.2\%) exhibiting outside-in quenching, 40 (25.6\% $\pm$ 3.5\%) showing inside-out quenching, one (0.6\% $\pm$ 0.6\%) showing globally quenching, five (3.2\% $\pm$ 1.4\%) being quenched overall, and 52 (33.3\% $\pm$ 3.8\%) with no clear trend. Specifically, the four CPSB galaxies are classified into three outside-in and one globally quenching. The seven RPSB galaxies are classified as five outside-in quenching, one being AGN contaminated and one with no clear trend. For the 145 IPSB galaxies, 27 are AGN contaminated, 22 exhibit outside-in quenching, 40 inside-out, five quenched overall, and 51 with no clear trend. The fractions of these six quenching classes and the factors of excesses in post-mergers relative to controls are summarized in Table \ref{tab:f_quench}.

As seen in Table~\ref{tab:f_quench}, outside-in quenching is more common in post-mergers than in control galaxies, with an excess of a factor of 2.9 $\pm$ 0.6. 
Galaxies which are globally quenching or have been fully quenched all over the entire galaxy have similar fractions in post-mergers and control galaxies considering the errors. Galaxies with inside-out quenching or quenching without a clear trend tend to be more common in non-merging systems than in post-mergers.
In addition, outside-in quenching is $\sim$7 times more common than inside-out quenching in post-mergers while inside-out quenching is slightly more common in non-merging controls, suggesting mergers leading to outside-in quenching. 
This indicates that the mechanisms driving quenching in mergers are preferentially operating in the disk rather than in the nuclear center. It could be due to inflows or turbulence redistributing the gas. Gas consumption of trigger starbursts and stellar feedback in the disk can also play a role. On the contrary, merger-driven AGN feedback is not likely the main mechanism driving quenching in mergers. 
However, we cannot rule out that AGN feedback may have long-term effects on maintaining quenching rather than in short merger timescales as suggested in \cite{2022MNRAS.515.1430D}.

\begin{figure*}
\begin{center}
\begin{minipage}{0.32\textwidth}
\includegraphics[width=\linewidth]{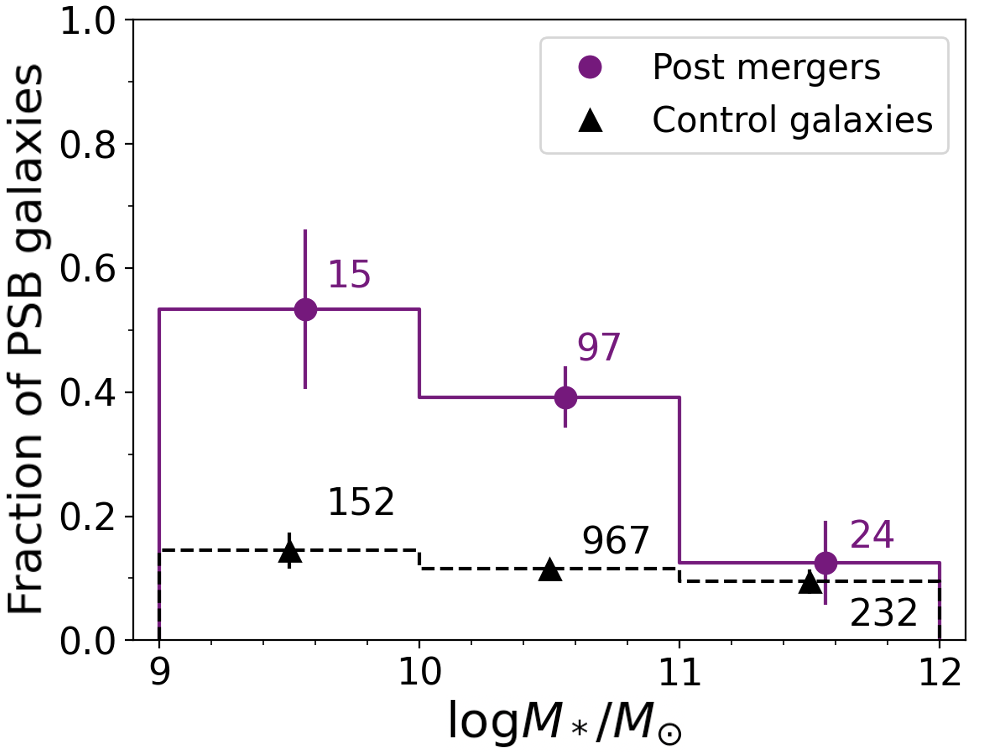}
\end{minipage}
\begin{minipage}{0.32\textwidth}
\includegraphics[width=\linewidth]{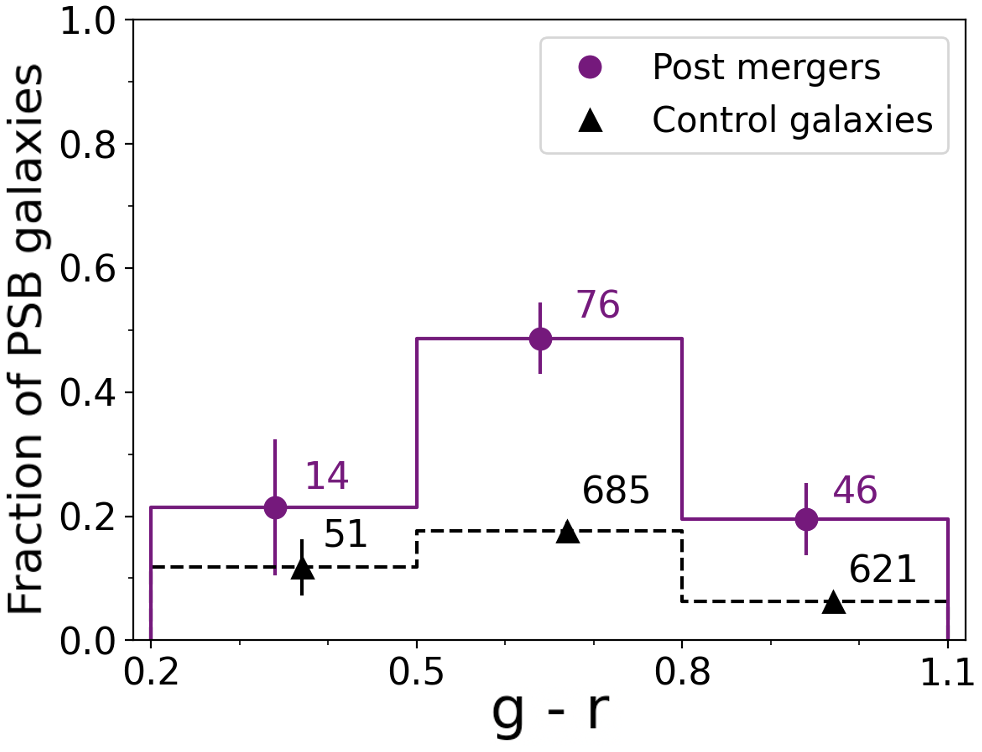}
\end{minipage}
\begin{minipage}{0.32\textwidth}
\includegraphics[width=\linewidth]{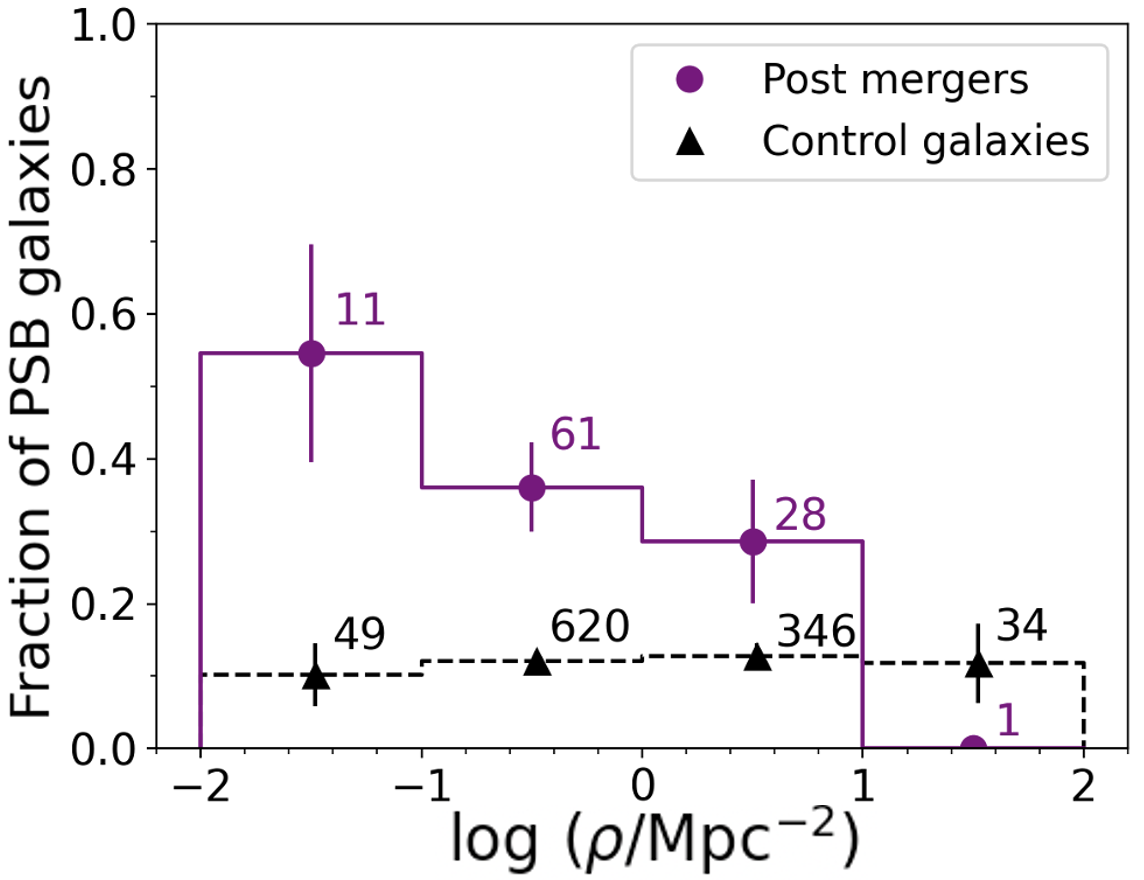}
\end{minipage}
\caption{Fraction of resolved PSB galaxies as a function of stellar mass, g - r color and environment density $\rho$ in MaNGA observed post-merger (solid purple) and control galaxies (black dashed). The PSB galaxies include all the CPSBs + RPSBs + IPSBs classified in MaNGA resolved spectra. The numbers of galaxies in each bin are shown next to the error bars. Post-mergers have a higher fraction of PSBs than control galaxies generally. In addition, PSB galaxies tend to appear more in galaxies with lower stellar mass, intermediate color and lower environment density.}
\label{fig:psb_galaxyproperties}
\vspace{-0.2cm}
\end{center}
\end{figure*}

Once we restrict our samples to only the CPSBs and RPSBs (see Table~\ref{tab:f_quench_CR}), there is no difference found in each of the quenching directions in post-mergers compared to controls. However, we found that outside-in quenching is still much more common than inside-out quenching in post-mergers. In fact, the CPSBs and RPSBs mainly show outside-in quenching regardless of post-mergers or controls. These results suggest that outside-in quenching is playing the main role in triggering CPSBs and RPSBs in both post-mergers and control galaxies.

\subsection{Resolved PSB Frequency vs Galaxy Properties}
To investigate whether PSB galaxies have a dependence on galaxy properties, in Figure~\ref{fig:psb_galaxyproperties} we compared the fractions of resolved PSB galaxies as a function of stellar mass (left panel), g - r color (middle panel), and environment density (right panel) in the MaNGA observed post-merger and control samples. 
The environment densities of our galaxies were obtained from \cite{2006MNRAS.373..469B}, which are defined as: $\rho = N/(\pi d^2)$, where $d$ is the projected co-moving distance to the N-th nearest neighbour within a specific redshift range (with details in \citealp{2006MNRAS.373..469B}). 
It should be noted that 100/136 (73.5\%) post-mergers and 519/1,360 (38.2\%) control galaxies in the MaNGA observed sub-sample have environment density measurements. However, we do not have the same number of control galaxies for each post-merger in this sub-sample, which may induce some bias.

We found the fraction of PSB galaxies decreases with increasing stellar mass in post-mergers. However, the fraction of PSB galaxies is independent of stellar mass in controls given our measurement uncertainties. It indicates that quenching in mergers is more common in low-mass systems. This is consistent with \cite{2018MNRAS.473.1168R} that there are more low-mass systems in PSB galaxies.
The middle panel shows that PSB galaxies tend to exhibit intermediate colors (g - r $\sim$ 0.5 -- 0.8) for both post-mergers and control galaxies. This is consistent with the expectation that the PSB phase is more likely to be found in the transition phase (or `green valley') between star-forming and quiescent. Redder galaxies in our sample tend to have lower fractions of PSBs as they are more likely to be quenched systems. 
From the right panel, the PSB fractions in post-merger galaxies tends to be higher at lower environment densities. PSB fractions in control galaxies show no clear dependence on environment density. However, it should be noted that our sample does not probe large clusters. 
In sum, star formation quenching in post-mergers tend to appear more frequently in lower mass, intermediate color, and lower environment density. Quenching in control galaxies is more frequently found in blue-cloud or green-valley galaxies while there is no trend found in stellar mass or environment density.
We also investigated the distribution of stellar mass, color, and environment density of the three different types of PSB (CPSB, RPSB, IPSB) galaxies. However, no clear trend is found in those distributions due to the small sample size.

\subsection{HI Gas Content in Resolved PSB Galaxies}
As seen in Figure~\ref{fig:psb_galaxyproperties}, PSB galaxies are more common in systems with lower mass and intermediate colors. As both color and stellar mass have a tight relation to the gas content in galaxies, we expect to see a difference between the gas content in PSBs compared to non-PSBs. We obtained the HI gas mass from the MaNGA-HI catalog and calculated the HI gas fraction, which is defined as $f_{gas}$ = M$_{HI}$/M$_{*}$.

\begin{figure}
\begin{center}
\begin{minipage}{0.47\textwidth}
\includegraphics[width=\linewidth]{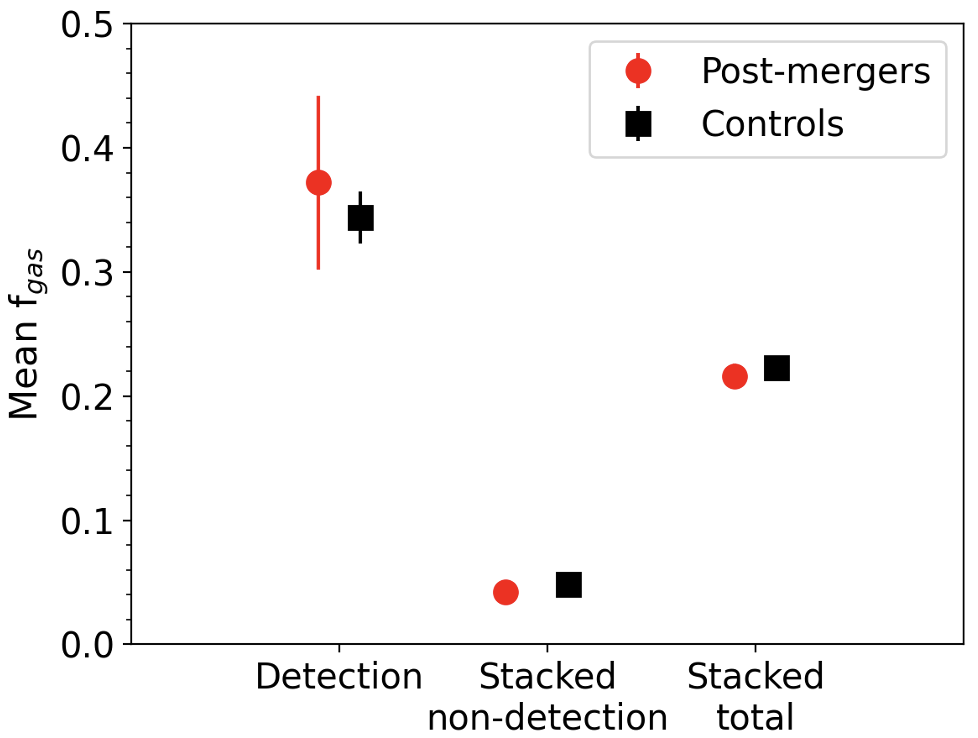}
\end{minipage}
\caption{The mean HI gas fraction in post-mergers (red circle) and control galaxies (black square). For the 37 post-mergers and 330 controls with HI detections, we calculated the mean $f_{gas}$ of the sample with standard 1$\sigma$ error. For the 42 non-detections in post-mergers and 563 in controls, we obtained the mean $f_{gas}$ and error by stacking their HI spectra. The mean $f_{gas}$ and error of the entire post-merger or control sample was obtained by stacking both detections and non-detections. }
\label{fig:fgas_pm_control}
\end{center}
\end{figure}

\begin{figure*}
\begin{center}
\begin{minipage}{0.98\textwidth}
\includegraphics[width=\linewidth]{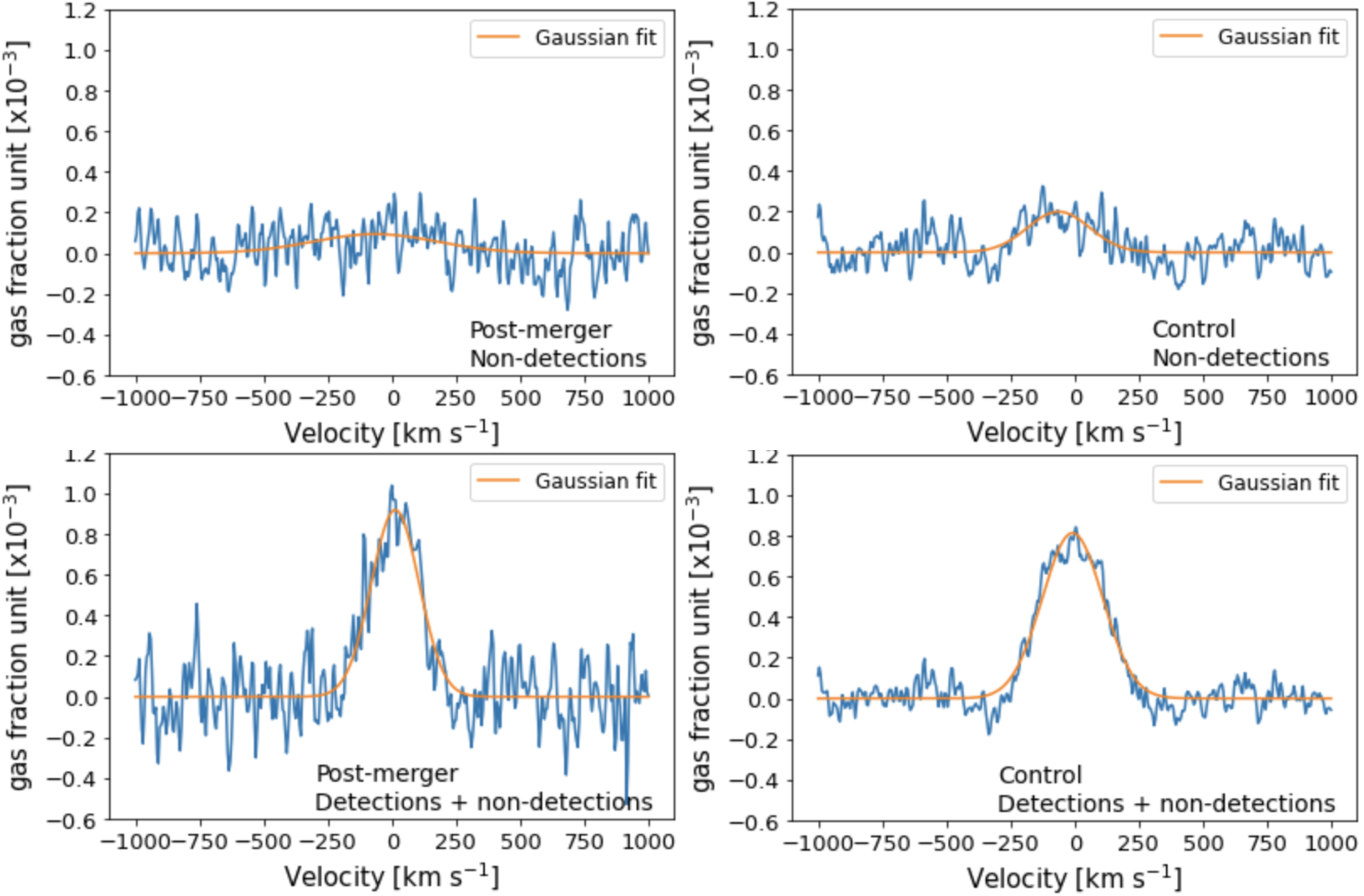}
\end{minipage}
\caption{The stacked HI spectra of non-detections on the upper panel and of detections + non-detections on the lower panel for post-mergers on the left and controls on the right. The spectra are centered at a velocity grid ranging between +/- 1000 km $s^{-1}$ and are shown in gas fraction unit ($M_{HI}/M_*$) converted from HI flux using equation~\ref{eq:1}.}
\label{fig:stack}
\end{center}
\end{figure*}

There are 37 post-mergers and 330 control galaxies with HI detections from the MaNGA-HI DR3 catalog. The mean $f_{gas}$ with 1$\sigma$ error is 0.372 $\pm$ 0.070 in post-mergers and 0.344 $\pm$ 0.021 in controls.
For the 42 post-mergers and 563 controls with HI non-detections, we obtained their mean $f_{gas}$ through stacking analysis following \cite{2020MNRAS.493.3081R}.
To summarize our stacking procedure: 
(1) We scaled all the HI spectra to the rest frame and converted them from flux density units to gas fraction units by using the following equation:
\begin{equation}\label{eq:1}
M_{HI}/M_* = 2.356\times10^5 (d[Mpc])^2 \frac{flux[Jy]}{M_*}
\end{equation}
where $d$ is the distance to the galaxy calculated using $d=cz/H_0$, where $c$ is the speed of light, $z$ is the redshift of the galaxy and $H_0=70$km s$^{-1}$Mpc$^{-1}$. Adopting the distance factor eliminates the bias of HI flux towards nearer galaxies. 
(2) We then recentered all the spectra onto a new velocity grid ranging between $\pm$ 1000 km s$^{-1}$ with interpolation at 5 km s$^{-1}$ intervals and added them to create the stack.
(3) Lastly, we averaged the stack to obtain the final average stacked spectra. 
We used a polynomial model to fit the baseline and integrated the baseline-subtracted profile using a Gaussian fit to obtain the mean $f_{gas}$ of the stack.

The mean stacked $f_{gas}$ of the non-detections is 0.042 $\pm$ 0.007 in post-mergers and 0.048 $\pm$ 0.005 in controls. 
To obtain the mean $f_{gas}$ for the entire post-merger sample and the control sample, we stacked both the detections and non-detections using the same procedure above. 
Figure~\ref{fig:stack} shows the stacked HI spectra of non-detections on the upper panel and of detection + non-detections on the lower panel for post-mergers on the left and control galaxies on the right. 
Stacking both detections and non-detections, we obtained a mean stacked total $f_{gas}$ of 0.216 $\pm$ 0.010 in post-mergers and 0.223 $\pm$ 0.004 in controls. 
As can be seen in Figure~\ref{fig:fgas_pm_control}, there is no difference between the mean $f_{gas}$ in post-mergers and controls within 1$\sigma$ error for either the detected or stacked sub-samples.

It should be noted that \cite{2018MNRAS.478.3447E} studied a sample of 98 post-mergers and found a median HI gas enhancement of $\sim$ 0.51 dex in post-mergers relative to controls matched at stellar masses, suggesting that gas exhaustion is not the cause of star formation quenching in mergers. Our MaNGA sample of 79 post-mergers is comparable in size to \cite{2018MNRAS.478.3447E}, although our sample has more non-detections. 
A control sample without being matched at redshift may cause a difference in the gas fraction enhancement.
In addition, we used a different method to compare the $f_{gas}$, which may also affect the results.
Sample biases such as a dependence on stellar mass, redshift, environment cannot be ruled out with the relatively small sample. In a future paper we will investigate the dependence of gas fraction offset with a much larger sample of post-mergers and non-interacting controls. 

\begin{figure}
\begin{center}
\begin{minipage}{0.47\textwidth}
\includegraphics[width=\linewidth]{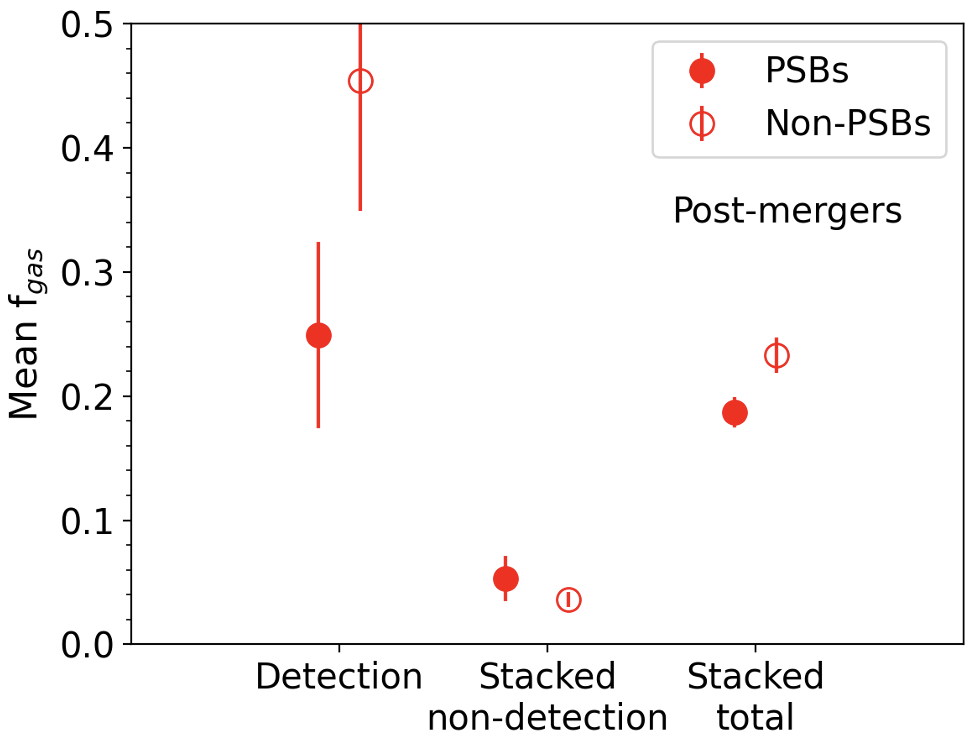}
\end{minipage}
\begin{minipage}{0.47\textwidth}
\includegraphics[width=\linewidth]{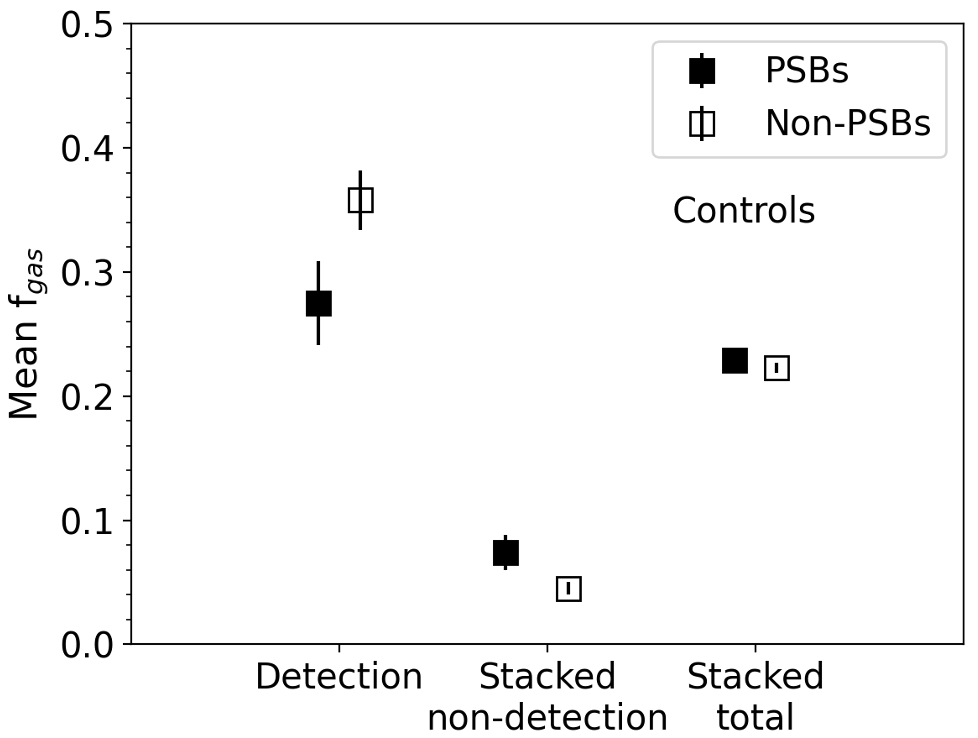}
\end{minipage}
\caption{Upper panel: The mean HI gas fraction in PSBs (filled circles) and non-PSBs (empty circles) in post-mergers. There are 15 PSBs and 22 non-PSBs with HI detections while 13 PSBs and 29 non-PSBs have HI non-detections. There is no detection in the stack of the 29 non-detections. Hence, the upper limit is shown. PSBs tend to have less gas than non-PSBs in post-mergers. 
Lower panel: the mean HI gas fraction in PSBs (filled squares) and non-PSBs (empty squares) in control galaxies. There are 55 PSBs and 275 non-PSBs with HI detections while 50 PSBs and 513 non-PSBs have HI non-detections. There is no difference in the gas fraction of PSBs compared to non-PSBs in controls.}
\label{fig:fgas_psb_pm_control}
\end{center}
\end{figure}

In Figure~\ref{fig:fgas_psb_pm_control}, we investigate the dependence of mean gas fraction on the presence/absence of resolved PSB (CPSBs, RPSBs, IPSBs) features in post-mergers and controls.
In post-mergers, there are 15 PSBs and 22 non-PSBs with HI detections. The mean $f_{gas}$ is 0.249 $\pm$ 0.075 in PSBs and 0.454 $\pm$ 0.105 in non-PSBs. 
For the non-detections, the mean $f_{gas}$ of the 13 PSBs is 0.053 $\pm$ 0.018 while the mean $f_{gas}$ of the 29 non-PSBs is 0.036 $\pm$ 0.006.
By stacking both detections and non-detections, we obtained a mean stacked total $f_{gas}$ of 0.187 $\pm$ 0.012 in PSBs and 0.233 $\pm$ 0.014 in non-PSBs.
Figure~\ref{fig:fgas_psb_pm_control} upper panel shows the mean $f_{gas}$ in PSBs and non-PSBs in post-mergers. The detections show a slight gas deficit in PSBs relative to non-PSBs at a 2$\sigma$ level. The stacked non-detections show no difference in the mean $f_{gas}$ in PSBs compared to non-PSBs. Considering both detections and non-detections, there is a small deficit of the gas fraction of 5\% in PSBs compared to non-PSBs in post-mergers. 
This suggests gas consumption/expulsion may play a role in triggering PSB signatures in post-mergers.

For control galaxies, there are 55 PSBs and 275 non-PSBs with HI detections. The mean $f_{gas}$ is 0.275 $\pm$ 0.034 in PSBs and 0.358 $\pm$ 0.024 in non-PSBs. For non-detections, the mean $f_{gas}$ is 0.074 $\pm$ 0.014 in the 50 PSBs and 0.045 $\pm$ 0.005 in the 513 non-PSBs. Stacking both detections and non-detections, the mean $f_{gas}$ is 0.229 $\pm$ 0.008 in PSBs and 0.223 $\pm$ 0.004 in non-PSBs. 
Similar to post-mergers, the detections show a slight gas deficit in PSBs relative to non-PSBs. The stacked non-detections show a slight gas enhancement in control PSBs. However, accounting for both detections and non-detections, there is no difference between the $f_{gas}$ in PSBs and non-PSBs in controls. This suggests gas consumption/expulsion is not playing a role in triggering quenching in non-merging galaxies.

To summarize, we found no difference in the gas fraction between post-mergers and controls. However, we found post-mergers exhibiting PSB features have an HI gas deficit compared to post-mergers not hosting PSB signatures. Control galaxies show no such trend. This is consistent with gas consumption/star-formation driven expulsion as being the cause of the PSB signatures in mergers. 
It should be noted that considering only CPSBs + RPSBs as PSB galaxies does not change our results.

\section{Discussion}
\subsection{Post-Starburst Excess in Post-Merger Galaxies}
Here we compare the PSB excesses found in our post-merger sample to a similar work by \cite{2022MNRAS.517L..92E}, which investigated the PSB fraction in a sample (E22 sample, hereafter) of 508 post-merger galaxies with log M$_*$/M$_{\odot} >$ 10 and z $<$ 0.25. By using the \cite{2007MNRAS.381..187G} traditional `E+A' PSB classification method, they found 6\% $\pm$ 1\% of post-mergers and 0.1\% $\pm$ 0.02\% of non-merging controls classified as PSBs, with a PSB excess of a factor of 60 $\pm$ 16 in post-mergers. In addition, a Principal Component Analysis (PCA) analysis was also presented in their work to classify PSBs, which results in a PSB excess of a factor of 33 $\pm$ 4 in post-mergers. 

The PSB excesses in the E22 sample are higher than those of our post-merger samples with single-fiber data (9.9$~\sim~$20). One of the reasons for these differences can be the different stellar mass and redshift range of our sample. The post-merger sample in our work spans a wider range of stellar mass (9 $<$ log M$_*$/M$_{\odot} <$ 12) and a much narrower range of redshift (0.02 $<$ z $<$ 0.06) compared to the E22 sample. 
If we restrict our DR14 post-merger sample to the same stellar mass range as the E22 sample, we will obtain 914 post-mergers. When using the \cite{2007MNRAS.381..187G} traditional `E+A' PSB classification method, the number statistics are too small to draw a robust conclusion. 
We also checked the PCA PSB fractions in our mass-restricted DR14 post-merger sample by using the catalog presented in \citep{2007MNRAS.381..543W}. We found a PCA PSB excess of a factor of 22.8 $\pm$ 5.4 in our post-mergers, which is slightly lower than the PCA PSB fraction of the E22 sample (33 $\pm$ 4).
However, when using the \cite{2019MNRAS.489.5709C} single-fiber classification, we obtained a PSB excess of a factor of 23.1 $\pm$ 6.9, which is consistent with the PCA PSB excess of our post-merger sample or the E22 sample. 
If we restrict the E22 sample to the same redshift and stellar mass range as our DR14 post-merger sample then the sample size is too small to draw any robust conclusion.

Our work here shows that single-fiber diagnostics can only identify CPSBs and strongly underestimates the total number of PSB galaxies. 
As seen in Table~\ref{tab:f_psb} and \ref{tab:cri_psb}, the resolved PSB diagnostic based on MaNGA IFU spectra reveals a large number of RPSBs and IPSBs showing quenching signatures in the outer disk (up to $\sim$2.5 $R_e$) rather than only in the center (CPSBs), which are missed by the single-fiber diagnostics. 
Specifically, $\sim$62\% of the resolved PSBs in post-mergers and $\sim$64\% in controls are missed by the single-fiber selection method when considering both CPSBs and RPSBs. Including the IPSBs, the single-fiber PSB method will miss $\sim$84\% of the resolved PSBs in post-mergers and $\sim$97\% in controls.
However, it should be noted that some of the IPSBs showing sporadic PSB regions could be due to stochastic star formation decay rather than permanent quenching.

\begin{figure*}
\begin{center}
\begin{minipage}{0.47\textwidth}
\includegraphics[width=\linewidth]{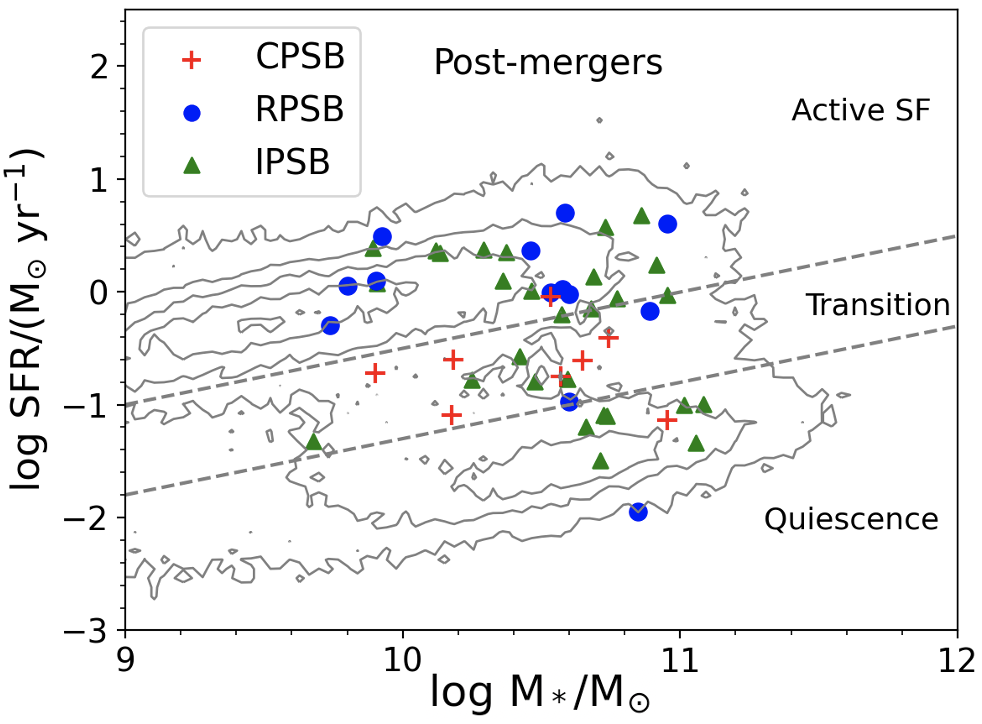}
\end{minipage}
\begin{minipage}{0.47\textwidth}
\includegraphics[width=\linewidth]{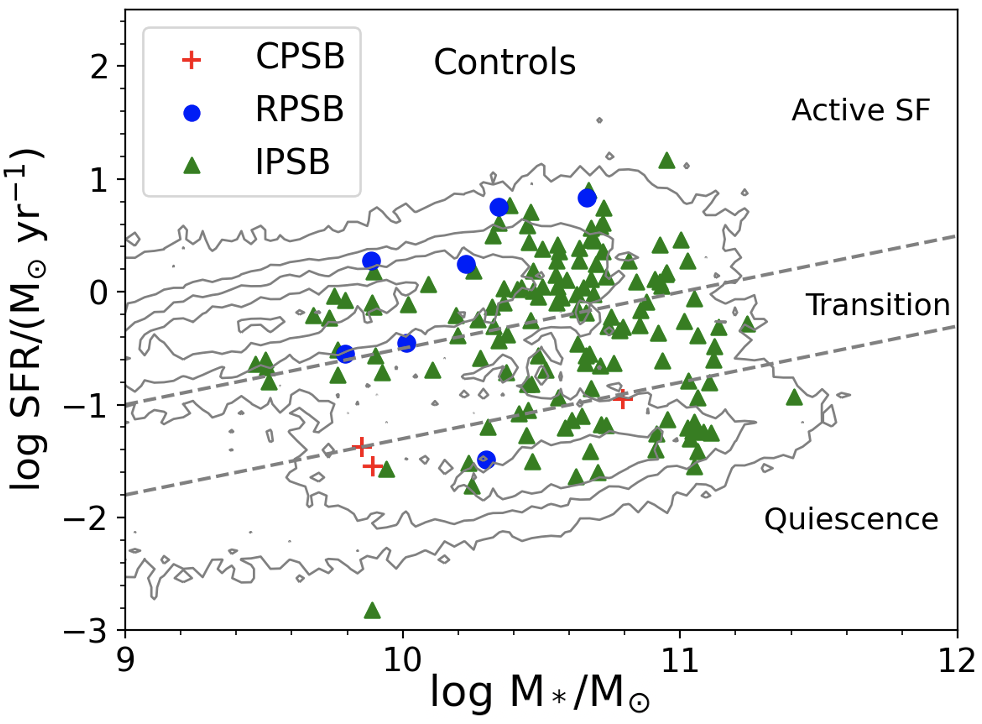}
\end{minipage}
\caption{Distribution of CPSBs (red crosses), RPSBs (blue circles) and IPSBs (green triangles) on the SFR vs. stellar mass diagram of post-mergers and controls. The contour represents the parent sample with the outer most contour including 99\% of the sample. The two division lines are drawn by eye based on the contour, which divide galaxies into active star-forming, transition and quiescence. RPSBs are mostly located in the active star-forming region. CPSBs are located along the transition region and the IPSBs span a range from star-forming to quiescent.}
\label{fig:ssfr_m}
\end{center}
\end{figure*}

\subsection{Quenching Mechanisms}
As seen in Table \ref{tab:f_quench}, the fraction of outside-in quenching in resolved PSB post-mergers is 44.9\% $\pm$ 7.1\%, about 7 times higher than the fraction of inside-out quenching (6.1\% $\pm$ 3.4\%) while inside-out quenching is slightly more common in control galaxies than outside-in. 
This suggests that the star formation quenching in post-mergers is more likely to be related to processes happening in the disk (i.e., gas inflows/consumption during mergers) rather than those happening in the center (merger-induced AGN/star formation). 
The HI gas deficit seen in post-merger PSBs relative to non-PSBs (see Figure~\ref{fig:fgas_psb_pm_control}) suggests that gas consumption from triggered star formation in the disk and the related stellar feedback may be the cause of PSBs and star formation quenching. 

Although our results are quite suggestive, previous studies \citep{2002ApJ...569..157W, 2007ApJ...671..333K, 2008AJ....136.2782L, 2008AJ....136.2846B} have found that star formation rates are better correlated with molecular gas but poorly related to atomic gas. 
The HI gas deficit may suggest either an efficient conversion of atomic Hydrogen to molecular Hydrogen \citep{2020ApJ...890...63W} and hence no total gas deficit or indicate a true gas deficit. Resolved distributions of dense molecular gas will be needed to truly constrain the mechanisms in play.

Figure \ref{fig:ssfr_m} shows the distribution of CPSBs, RPSBs and IPSBs on the SFR vs. stellar mass parameter space. In post-mergers, we find nearly all the RPSBs are located in the actively star-forming region while CPSBs are mostly located along the transition region. Although the IPSBs span a range from star-forming to quiescent, $\sim$75\% of the outside-in quenching IPSBs are located in the star-forming region. All of the inside-out or quenched overall IPSBs are located in the quiescent region. 
These results are consistent with \cite{2019MNRAS.489.5709C}, which found RPSBs are mainly located on the star forming main sequence while CPSBs are mainly located in the green valley. \cite{2019MNRAS.489.5709C} also found that CPSBs and RPSBs are mainly showing outside-in quenching scenarios.
This outside-in quenching scenario can be possibly due to strong gas inflows during mergers redistributing gas from the outer regions to the center and triggering strong central starbursts. While the post-starburst in the outer regions fades and becomes quiescent, the central starburst is quenched due to gas consumption. In addition, a merger-driven starburst progressively moving inwards followed by a truncation of star formation from the outside-in can also cause this quenching scenario.

In the resolved control PSBs (CPSBs + RPSBs + IPSBs), the fraction of inside-out quenching (25.6\% $\pm$ 3.5\%) is similar to the fraction of outside-in quenching (19.2\% $\pm$ 3.2\%). It seems neither mechanism is playing a dominant role in quenching star formation in non-interacting control galaxies. 
As seen in Figure~\ref{fig:ssfr_m} right panel, six out of the seven RPSBs in controls are located in the star-forming region while the other one is in the quiescent region.
Three out of the four CPSBs are located along or close to the transition region while the other one does not have valid measurements on SFR.
The distribution of RPSBs and CPSBs in controls is similar to that in post-mergers. IPSBs with outside-in quenching are located in the star-forming region while those showing inside-out or quenched overall are in the quiescent region. This is also similar to the IPSBs in post-mergers.

\subsubsection{The Role of AGN}
AGN have been found in PSB galaxies by many previous studies \citep{2006MNRAS.369.1765G, 2007ApJ...663L..77T, 2015A&A...582A..37M, 2017MNRAS.470.1687B, 2018MNRAS.480.3993B} and are believed to play a role in quenching star formation by feedback processes. However, other studies suggest that the AGN-driven outflows may not be able to remove a significant amount of gas to quench star formation globally \citep{2017arXiv170609500Y}, or AGN are just along for the ride rather than the primary cause of PSB signatures \citep{2014ApJ...792...84Y, 2019MNRAS.489.1139M, 2020ApJ...900..107Y, 2022ApJ...935...29L, 2022MNRAS.517L..92E}. 

As discussed in the appendix, we identified AGN in our samples using both single-fiber and resolved optical diagnostics. We found that the optical AGN fraction in PSBs is a factor of $\sim$ two times higher than that in non-PSBs in our samples. This is true in both post-mergers and control galaxies, suggesting a possible relation between AGN and PSBs. 
However, the existence of AGN is not enough to correlate the cause of PSBs and star formation quenching to AGN feedback processes. This is because the mechanisms (e.g., galaxy mergers) which trigger the PSBs and star formation quenching could be the same mechanisms triggering the AGN \citep{2022MNRAS.517L..92E}. Hence, the AGN could be the by-product of the real triggering mechanism of PSB.
As seen in Table~\ref{tab:f_quench}, we found a consistent fraction of resolved PSBs hosting an AGN in post-mergers ($\sim$22\%) and in control galaxies ($\sim$18\%), which prevents us from determining their resolved quenching direction. This suggests merger-driven AGN is not the main quenching mechanism in post-mergers.
We also found that the AGN fraction in different types of PSB galaxies (CPSBs, RPSBs and IPSBs) are similar to each other in post-mergers, suggesting AGN are not the cause of the variety of resolved PSB signatures in mergers. 

AGN can be identified in multi-wavebands, and different AGN identification methods may reveal very different populations (see \citealp{2017A&ARv..25....2P} for a review). Using the optical diagnostic alone cannot provide a complete census of the AGN population. A recent paper by \cite{2023ApJ...944..168L} studies the AGN fraction using multi-waveband diagnostics in the same post-merger sample as in this work. They showed that X-ray diagnostics are crucial in identifying the majority of AGN in our post-merger sample. However, only 12 out of the 136 MaNGA observed post-mergers have deep X-ray observations. Hence, we cannot completely rule out a role for AGN in star formation quenching in post-mergers. 
In addition, our systems are visually identified post-mergers. Merger remnants with strong quasars would not have been visually identified and are therefore not included in this analysis. 

\cite{2022MNRAS.515.1430D} suggests that AGN feedback can cause quenching over a long timescale of a few Gyrs by expelling gas from the circumgalactic medium and halting further gas supply. This is similar to strangulation \citep{2015Natur.521..192P, 2020MNRAS.491.5406T}, which is another long-term quenching mechanism. However, the typical timescale of PSB signatures is about a Gyr after the starburst, which is shorter than the timescales for strangulation (3$\sim$4 Gyr). For galaxies with gradual quenching over long timescales, their spectra are not expected to exhibit both strong Balmer absorption and no emission lines (i.e., PSB).

In addition to galaxy mergers and AGN, previous studies also found that ram pressure stripping plays a role in triggering PSB galaxies in dense clusters \citep{2009ApJ...693..112P, 2013ApJ...770...62D, 2017ApJ...838..148P, 2018MNRAS.476.1242S, 2019MNRAS.482..881P, 2021MNRAS.504.4533W, 2022ApJ...930...43W}. However, the post-merger galaxies in our sample are mainly in the field rather than in high density environments and hence ram pressure is unlikely to be playing a role in triggering PSBs in our sample.

\section{Summary}
In this paper, we investigated the role of galaxy mergers in triggering star formation quenching using a sample of 1,051 post-merger galaxies and 10,510 non-merging control galaxies. We classified PSB galaxies using both single-fiber spectra and the MaNGA resolved spectra. We presented visual classifications of the resolved star formation quenching history of our PSB galaxies and further classified our galaxies into different quenching directions. Here we summarize our results:

(i) Using the SDSS single-fiber spectra with different PSB selection methods, we found a consistent (with errors) PSB excess of a factor of $\sim$10-20 in post-mergers relative to non-merging controls. This suggests galaxy mergers play an important role in triggering PSB signatures and star formation quenching. The single-fiber PSB fractions and excesses are summarized in Table~\ref{tab:f_psb}.

(ii) Using the MaNGA resolved spectra, we identified three morphological classes of resolved PSBs: central (C)PSBs, ring-like (R)PSBs, and irregular (I)PSBs. 
When only accounting for CPSBs + RPSBs, we found a PSBs excess of a factor of $\sim$19 in post-mergers relative to controls, which is consistent with the single-fiber PSB excess. However, we found that $\sim$62\% of the resolved PSBs (CPSBs + RPSBs) in post-mergers and $\sim$64\% in controls were missed by the single-fiber PSB selection method. Specifically, the resolved IFU spectra are crucial to recover the RPSBs while the SDSS single-fiber spectra can only identify the CPSBs in our sample.
Including the IPSBs, we obtained a PSB excess of a factor of $\sim$3 in post-mergers relative to controls.
Some IPSBs may be due to sporadic star formation decay rather than permanent quenching. The resolved PSB fractions and excesses are shown in Table~\ref{tab:cri_psb}.

(iii) In post-merger galaxies, outside-in quenching is $\sim$7 times more common than inside-out quenching. However, inside-out quenching is slightly more common than outside-in quenching in non-merging controls (see Table~\ref{tab:f_quench}). This suggests that mergers preferentially lead to quenching from the disk rather than from the nuclear center, suggesting merger-driven AGN feedback is not the main quenching mechanism. Gas inflows, gas consumption, and stellar feedback driven in mergers may play a more important role.

(iv) In post-merger galaxies, the mean HI gas fraction in PSBs is lower than that in non-PSBs (see Figure~\ref{fig:fgas_psb_pm_control}), suggesting gas consumption/SF driven expulsion is the likely cause of PSB and star formation quenching in mergers as opposed to gas inflows alone.

\section*{Acknowledgement}
We thank the referee for the comments that have improved the quality of this work.
Authors Li and Nair would like to acknowledge support by the National Science Foundation under Grant No. 1616547. The NSF grant enabled the visual classification of the parent sample of galaxies used in this paper. Any opinions, findings, and conclusions or recommendations expressed in this material are those of the author(s) and do not necessarily reflect the views of the National Science Foundation.

The Green Bank Observatory is a facility of the National Science Foundation operated under cooperative agreement by Associated Universities, Inc.
We would like to acknowledge Zachary Tu from the University of Washington, who provided the stacking code used in our HI analysis and provides assistance with the HI related problems.

This publication makes use of data from the Sloan Digital Sky Survey. Funding for the Sloan Digital Sky Survey has been provided by the Alfred P. Sloan Foundation, the U.S. Department of Energy Office of Science, and the Participating Institutions. SDSS acknowledges support and resources from the Center for High Performance Computing at the University of Utah. The SDSS website is www.sdss.org. 

SDSS-IV is managed by the Astrophysical Research Consortium for the 
Participating Institutions of the SDSS Collaboration including the 
Brazilian Participation Group, the Carnegie Institution for Science, 
Carnegie Mellon University, the Chilean Participation Group, the French Participation Group, Harvard-Smithsonian Center for Astrophysics, 
Instituto de Astrof\'isica de Canarias, The Johns Hopkins University, Kavli Institute for the Physics and Mathematics of the Universe (IPMU) / 
University of Tokyo, the Korean Participation Group, Lawrence Berkeley National Laboratory, 
Leibniz Institut f\"ur Astrophysik Potsdam (AIP),  
Max-Planck-Institut f\"ur Astronomie (MPIA Heidelberg), 
Max-Planck-Institut f\"ur Astrophysik (MPA Garching), 
Max-Planck-Institut f\"ur Extraterrestrische Physik (MPE), 
National Astronomical Observatories of China, New Mexico State University, 
New York University, University of Notre Dame, 
Observat\'ario Nacional / MCTI, The Ohio State University, 
Pennsylvania State University, Shanghai Astronomical Observatory, 
United Kingdom Participation Group,
Universidad Nacional Aut\'onoma de M\'exico, University of Arizona, 
University of Colorado Boulder, University of Oxford, University of Portsmouth, 
University of Utah, University of Virginia, University of Washington, University of Wisconsin, 
Vanderbilt University, and Yale University.

\section*{Data Availability}
Part of the parent catalog used to construct the samples in this paper will be available in Nair (2023; in preparation). The MaNGA observed post-merger sample used in this paper is publicly available as an online supplement to its original journal publication.
The MaNGA MPL-11 data products can be generated by the public using the raw data at \href{https://www.sdss4.org/dr17/manga/manga-data/data-access/}{this link} with DRP v3.1.1 and DAP v3.1.0.
The MaNGA-HI DR3 catalog used in this paper can be found at \href{https://www.sdss4.org/dr17/data_access/value-added-catalogs/?vac_id=hi-manga-dr3}{this link}.
Stellar masses, SFRs and measurements of line equivalent widths are all publicly from the MPA-JHU DR8 catalog at \href{https://www.sdss4.org/dr12/spectro/galaxy_mpajhu/}{this link}. 
The environment density used in this paper are accessible at \href{https://www.astro.ljmu.ac.uk/~ikb/research/bimodality-paperIV.html}{this link}.

\appendix
\section{AGN Identification}\label{agn_frequency}
As mentioned in Section~\ref{resolvedPSB}, the optical emission due to AGN will affect our classification of the resolved star formation history. Here we investigate the optical AGN fraction in our MaNGA observed samples using both single-fiber and resolved AGN diagnostics.

\begin{figure}
\begin{center}
\begin{minipage}{0.47\textwidth}
\includegraphics[width=\linewidth]{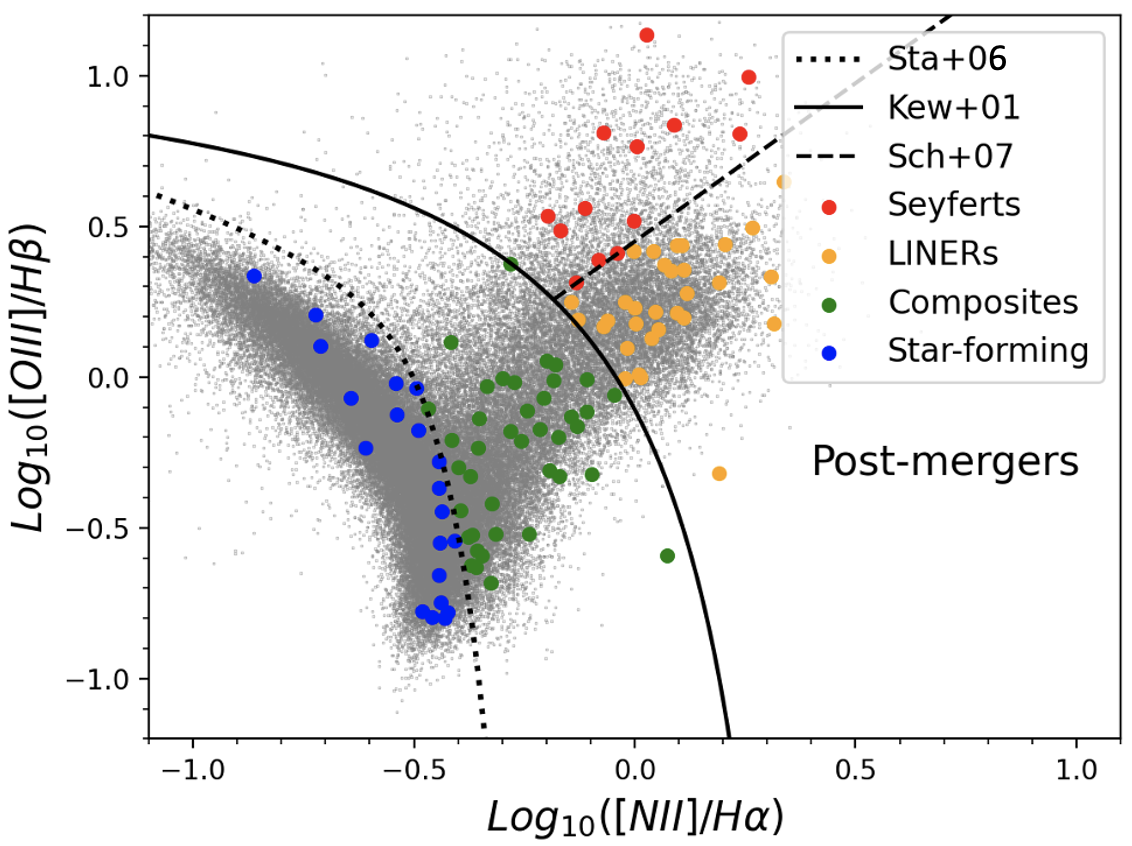}
\end{minipage}
\begin{minipage}{0.47\textwidth}
\includegraphics[width=\linewidth]{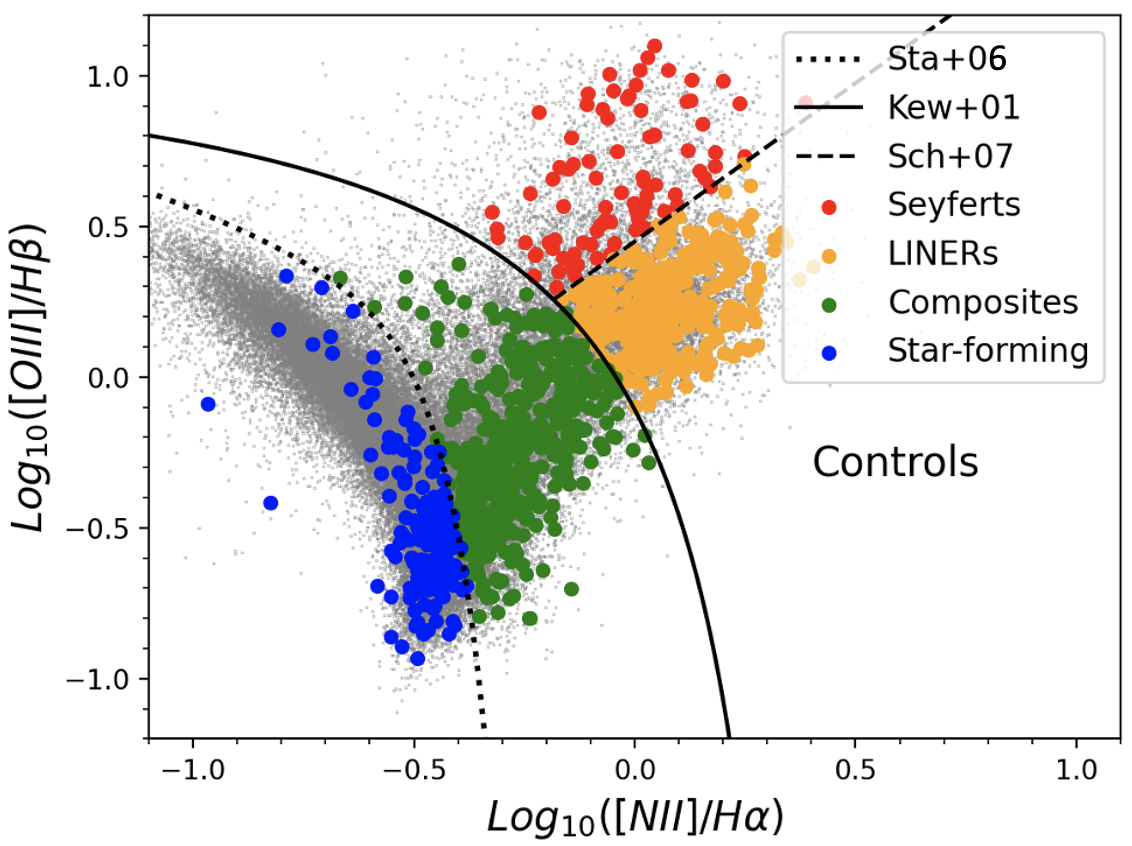}
\end{minipage}
\caption{The single-fiber BPT diagrams of post-mergers and control galaxies with S/N $\ge$ 3 for all four lines. The grey points are general galaxies from the parent sample. Galaxies are classified into Seyfert (red), LINERs (orange), composites (green) and star-forming (blue). Seyferts and LINERs are distinguished using the \citet{2007MNRAS.382.1415S} empirical (dashed) line. Objects between the \citet{Stasinska:2006uy} (dotted) and the \citet{2001ApJ...556..121K} (solid) curves are composites and those below the \citet{Stasinska:2006uy} curve are star-forming. Galaxies above the \citet{2001ApJ...556..121K} solid curve (Seyferts + LINERs) with EW(H$\alpha$) $\le$ -3\AA are classified as optical AGN.}
\label{fig:single_bpt}
\end{center}
\vspace{-0.5cm}
\end{figure}

In the single-fiber diagnostic, we used the emission line flux ratios of [O III]$\lambda5007$/H$\beta$ vs [N II]$\lambda6584$/H$\alpha$ measured from the SDSS single-fiber spectra to construct a NII-BPT diagram \citep{1981PASP...93....5B} with a cut on EW(H$\alpha$) $\le$ -3\AA\ \citep{2011MNRAS.413.1687C}. The fluxes of the four emission lines are obtained from the MPA-JHU catalog. We placed a S/N $\ge$ 3 cut on the four emission lines to select galaxies with significant flux measurements. Figure~\ref{fig:single_bpt} shows the single-fiber NII-BPT diagram of post-mergers and control galaxies. Galaxies with emission line S/N $<$ 3 are not shown on the diagrams. 
Out of the 136 post-mergers, the single-fiber diagnostic classifies 13 Seyferts, 31 LINERs, 40 composites, 21 star-forming and 31 low S/N.
Galaxies with emission line ratios above the \cite{2001ApJ...556..121K} curve (Seyferts + LINERs) and with EW(H$\alpha$) $\le$ -3\AA\ are classified as optical AGN. There are 16 optical AGN (eight Seyferts + eight LINERs) identified in post-mergers, implying an optical AGN fraction of 11.8\% $\pm$ 2.8\%. 
In the 1,360 control galaxies, there are 84 Seyferts, 351 LINERs, 423 composites, 155 star-forming and 347 low S/N classified by the single-fiber diagnostic.
Applying the equivalent width cut on EW(H$\alpha$), there are 108 optical AGN (58 Seyferts + 50 LINERs) identified in controls. This is an AGN fraction of 7.9\% $\pm$ 0.7\%. The AGN excess in post-mergers relative to controls is a factor of 1.5 $\pm$ 0.4.

\begin{figure*}[t]
\begin{center}
\begin{minipage}{0.98\textwidth}
\includegraphics[width=\linewidth]{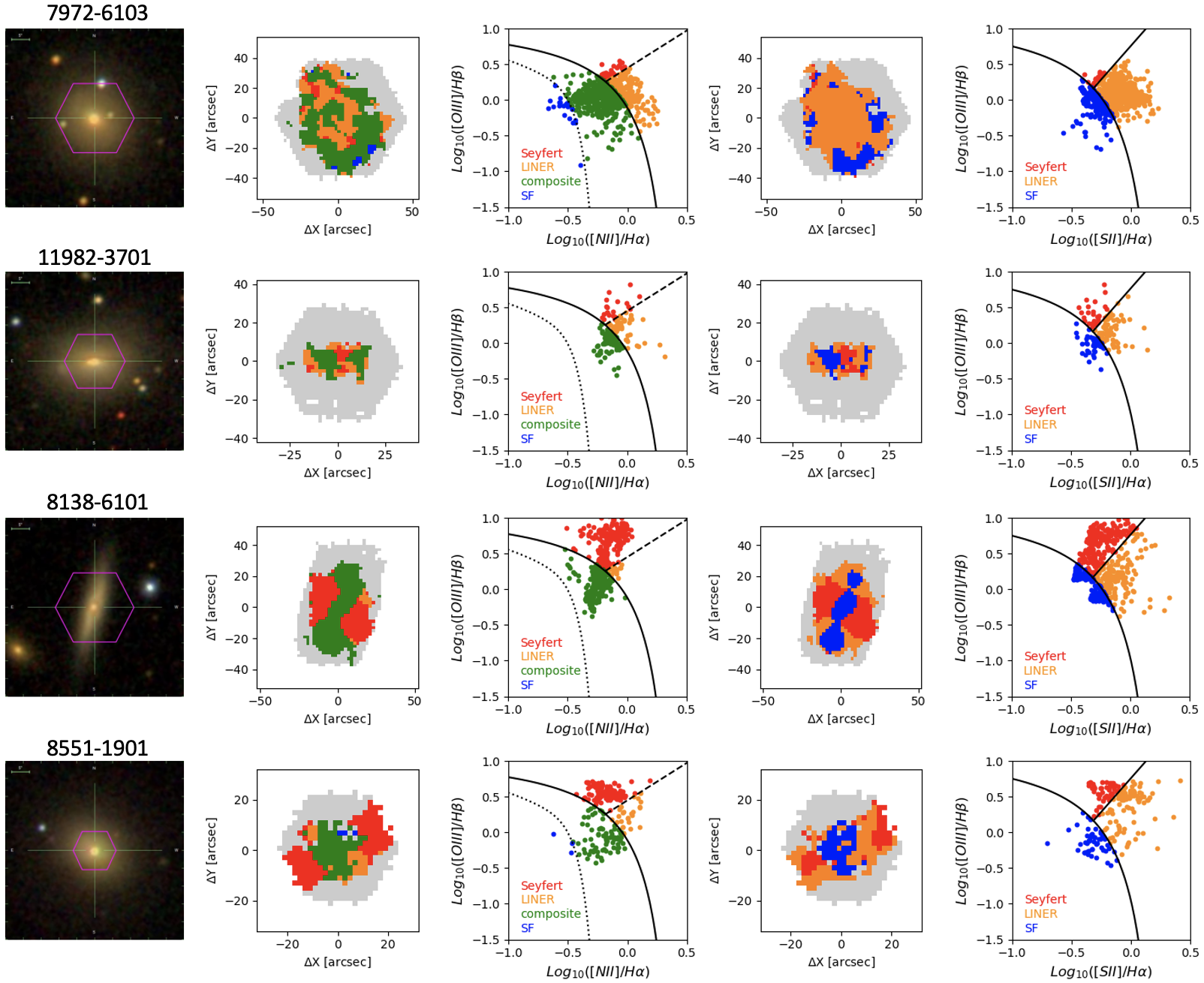}
\end{minipage}
\caption{Examples of resolved NII-BPT and SII-BPT diagrams for two post-mergers (top two panels) and two control galaxies (bottom two panels). Spaxels are classified into Seyfert (red), LINERs (orange), composites (green) and star-forming (blue). These four galaxies show clear resolved Seyfert signals but are classified as composites or low S/N by the single-fiber BPT diagnostic. Resolved diagnostic can reveal these optical AGN overlooked by the single-fiber spectra.}
\label{fig:bpt_cutout}
\end{center}
\end{figure*}

In the resolved diagnostic, we adopted and modified the criteria from \cite{2018MNRAS.474.1499W} by constructing two resolved BPT diagrams using the emission line flux ratios of [O III]/H$\beta$ vs [N II]/H$\alpha$ and [O III]/H$\beta$ vs [S II]/H$\alpha$ (NII-BPT, SII-BPT hereafter). The equivalent widths are extracted from the MaNGA ``MAPS-SPX-MILESHC-MASTARSSP'' data analysis pipeline. To obtain robust measurements, we only consider spaxels with a mean $g$-band spectral S/N $\ge$ 3 per pixel. We also required a S/N $\ge$ 3 cut on the fluxes of these five emission lines. As seen in Figure~\ref{fig:bpt_cutout}, spaxels are classified into Seyferts (red), LINERs (orange), composites (green) and star-forming (blue) by the \cite{2007MNRAS.382.1415S} (dashed) empirical relation, the \cite{2001ApJ...556..121K} (solid curved), \cite{2006MNRAS.372..961K} (solid, straight) and the \cite{Stasinska:2006uy} (dotted) theoretical relations. Spaxels with low S/N are shown in grey on the resolved maps and would not be shown on the BPT diagrams. 

We applied multiple cuts on the following parameters. Firstly, we required an AGN candidate to have at least 5\% of its spaxels classified as Seyferts or LINERs either by NII-BPT or SII-BPT. Secondly, as EW(H$\alpha$) is able to distinguish ionization mechanisms from real AGN and star formation \citep{2011MNRAS.413.1687C}, we required an AGN candidate to have a mean EW(H$\alpha$) of all the Seyfert + LINER spaxels to be less than -3\AA. Thirdly, diffuse ionized gas can affect the emission line ratio measurement particularly in the regions where the H$\alpha$ surface brightness is lower than 10$^{37}$ erg s$^{-1}$ kpc$^{-2}$ \citep{2017MNRAS.466.3217Z}. Following \cite{2018MNRAS.474.1499W}, we required an AGN candidate to have a mean surface brightness SB(H$\alpha$) of all Seyfert + LINER spaxels to be greater than 10$^{37.5}$ erg s$^{-1}$ kpc$^{-2}$. Lastly, Seyfert and LINER spaxels with emission line ratios close to the \cite{2001ApJ...556..121K} demarcation curve could be classified into composite (on NII-BPT) or star-forming (on SII-BPT) considering measurement errors. Composites on NII-BPT could possibly be AGN. However, star-forming spaxels misclassified as Seyferts or LINERs on SII-BPT should be excluded from real Seyfert and LINER spaxels. Following \cite{2018MNRAS.474.1499W}, we calculated the distance (d$_{BPT}$) from each Seyfert and LINER spaxel to the \cite{2001ApJ...556..121K} demarcation line on the SII-BPT diagram. We placed a cut on the mean d$_{BPT}$ of all Seyferts + LINERs spaxels and required an AGN candidate to have a mean d$_{BPT}$ $\ge$ 0.15 away from the star-forming demarcation line on the SII-BPT. It should be noted that our cut on d$_{BPT}$ is different from \cite{2018MNRAS.474.1499W}.

Here we summarize our AGN resolved classification methods. An AGN is classified if:
\begin{itemize}
    \item $f_{SL,NII} \ge 5\%$ 
    \item  mean EW(H$\alpha)_{SL,NII}\le-3$\AA 
    \item mean SB(H$\alpha)_{SL,NII} \ge 10^{37.5}$ erg s$^{-1}$ kpc$^{-2}$
\end{itemize}
OR
\begin{itemize}
    \item $f_{SL,SII} \ge 5\%$ 
    \item mean EW(H$\alpha)_{SL,SII}\le-3$\AA 
    \item mean SB(H$\alpha)_{SL,SII} \ge 10^{37.5}$ erg s$^{-1}$ kpc$^{-2}$ 
    \item mean d$_{BPT, SL,SII} \ge 0.15$
\end{itemize}
Using these resolved classification diagnostic, we identified 17 (12.5\% $\pm$ 2.8\%) AGN out of the 136 post-mergers and 74 (5.4\% $\pm$ 0.6\%) AGN out of the 1,360 control galaxies. It implies an AGN excess of a factor of 2.3 $\pm$ 0.6 in post-mergers.

\begin{figure*}
\begin{center}
\begin{minipage}{0.47\textwidth}
\includegraphics[width=\linewidth]{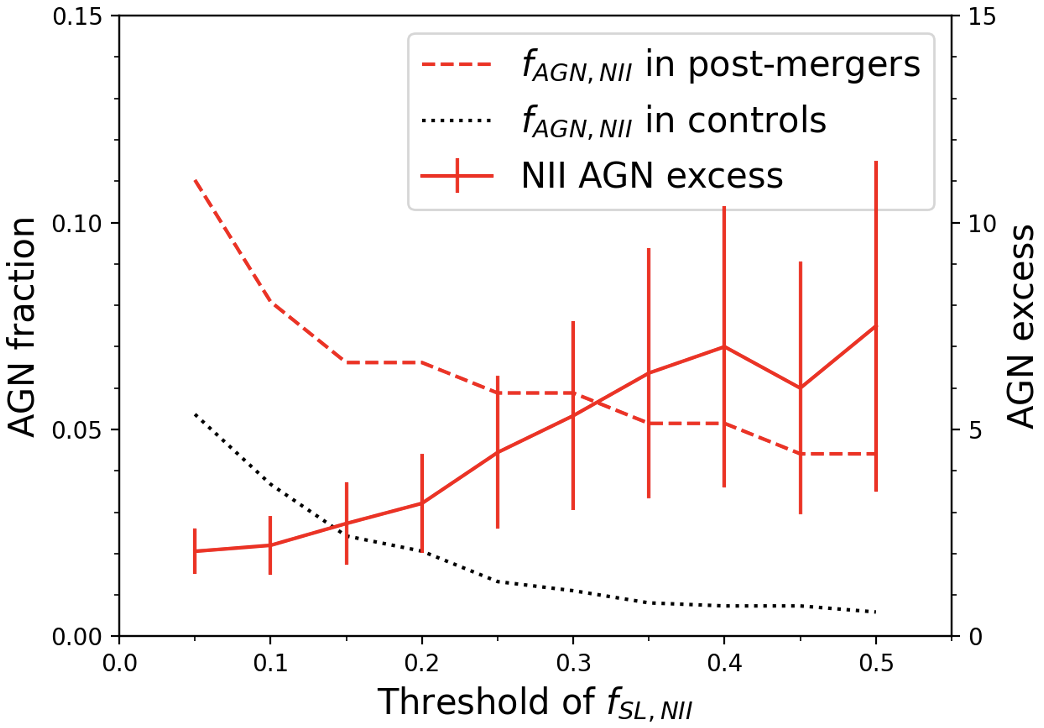}
\end{minipage}
\begin{minipage}{0.47\textwidth}
\includegraphics[width=\linewidth]{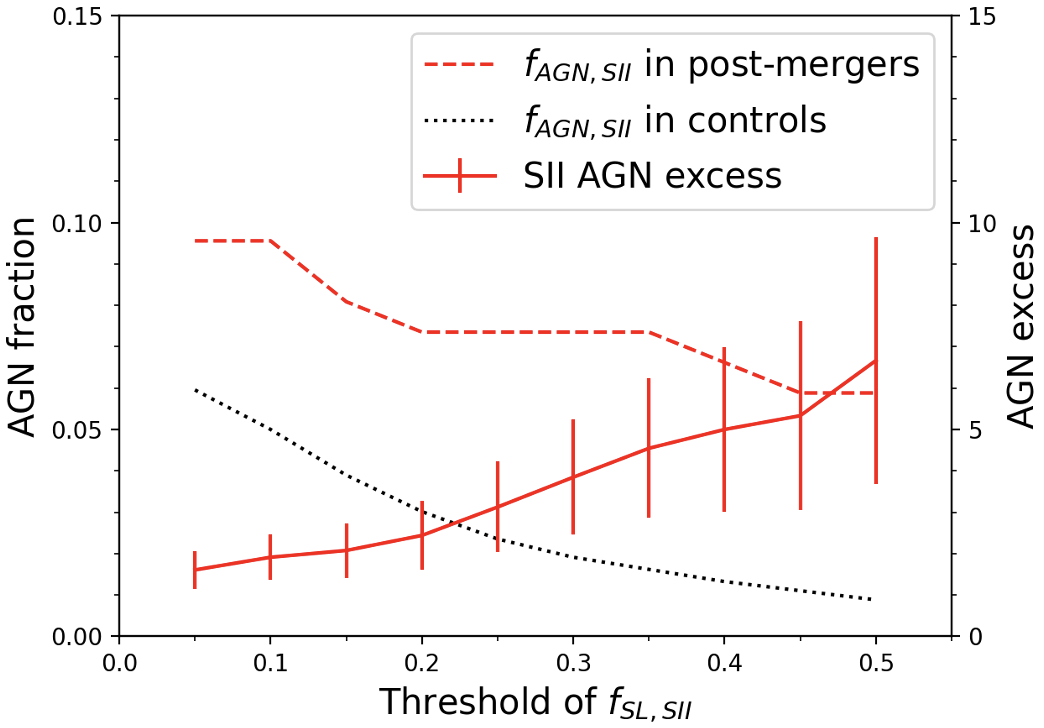}
\end{minipage}
\begin{minipage}{0.47\textwidth}
\includegraphics[width=\linewidth]{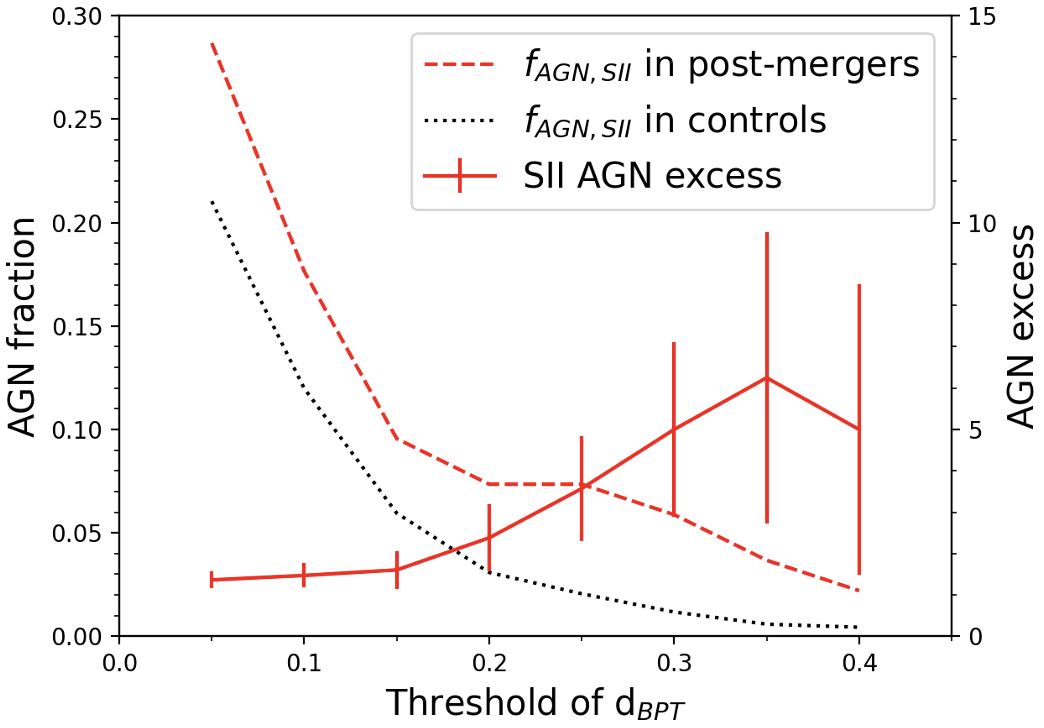}
\end{minipage}
\caption{AGN fraction in post-mergers (red dashed) and controls (black dotted) and AGN excess (red solid) with 1$\sigma$ errors as a function of Seyfert + LINER spaxel fraction ($f_{SL, NII}$, $f_{SL, SII}$) thresholds and BPT distance (d$_{BPT}$) thresholds. When investigating each threshold, the other cuts were set to the default values of our classification scheme. }
\label{fig:14}
\end{center}
\vspace{-0.4cm}
\end{figure*}

Unlike the cuts placed on the H$\alpha$ surface brightness \citep{2017MNRAS.466.3217Z} and EW(H$\alpha$) \citep{2011MNRAS.413.1687C} which are well supported by observations, the cuts we placed on the spaxel fractions and the d$_{BPT}$ are more empirical. Hence, we investigated how the AGN fractions and excesses are affected by choosing different thresholds. 
Figure~\ref{fig:14} shows the resolved AGN fractions and excesses in post-mergers relative to controls when using different thresholds of Seyferts + LINERs spaxel fractions ($f_{SL}$) and the distance away from the star-forming demarcation line (d$_{BPT}$). 
The top panel in Figure \ref{fig:14} shows the AGN identified in the NII-BPT and SII-BPT respectively. There is an AGN excess of a factor of 2 -- 8 in post-mergers relative to control galaxies at all considered thresholds. Choosing a higher threshold on the Seyferts + LINERs spaxel fraction will not qualitatively affect our results. 
In the bottom panel, the AGN fraction declines sharply at d$_{BPT}$ $<$ 0.15 and becomes smoother at d$_{BPT}$ $\ge$ 0.15. It suggests a small d$_{BPT}$ may not be robust enough to distinguish the contamination of star-forming spaxels in Seyfert/LINER spaxels and will include misclassified AGN in both post-merger and control samples. However, using a larger d$_{BPT}$ ($>$ 0.3) value will shrink the sample size and cannot lead to a robust conclusion of our results. In addition, there is an AGN excess in post-mergers relative to controls in all considered thresholds. 
In summary, using a higher threshold of spaxel fraction or d$_{BPT}$ does not affect our results qualitatively. Although it will slightly bring down the AGN fraction in both post-mergers and control galaxies and induce larger errors, the AGN excesses remain consistent considering errors.

In summary, the AGN fraction in post-mergers is 11.8\% $\pm$ 2.8\% based on single-fiber spectroscopy and 7.9\% $\pm$ 0.7\% for control galaxies. With resolved IFU data, we obtain and AGN fraction of 12.5\% $\pm$ 2.8\% for post-mergers and 5.4\% $\pm$ 0.6\% for control galaxies.
Considering AGN identified in either single-fiber BPT or resolved BPT diagnostics, the AGN fraction is 18.4\% $\pm$ 3.3\% in post-mergers and 10.0\% $\pm$ 0.8\% in controls. It implies an AGN excess of a factor of 1.8 $\pm$ 0.4 in post-mergers relative to controls.

\input{mnras_template.bbl}


\bsp	
\label{lastpage}
\end{document}